\documentclass[10pt,twocolumn,letterpaper]{article}

\usepackage[pagenumbers]{iccv} % To force page numbers, e.g. for an arXiv version
\usepackage{algorithmic}
\usepackage{algorithm}
\usepackage{xcolor} % For more color options

\definecolor{monokaibg}{HTML}{272822}

\usepackage[table,xcdraw]{xcolor}
\usepackage{bbm}
\newcommand{\myparagraph}[1]{ \vspace{3pt}  \noindent {\bf #1}\,\,\,}

\newcommand{\scenename}[1]{\textit{#1}}

\usepackage{amssymb,amsmath,stackengine}

\newcommand{\circumcenter}{\mathbb{O}}
\newcommand{\circumradius}{\mathbb{R}}
\newcommand{\centroid}{\boldsymbol{\mu}}
\newcommand{\ray}{\bold{r}}
\newcommand{\colorc}{\mathbf{c}}
\newcommand{\cgradient}{\nabla \colorc}
\newcommand{\sh}{\mathit{sh}}
\newcommand{\vertex}{\mathbf{v}}

\newcommand\lft{\mathopen{}\left}
\newcommand\rgt{\aftergroup\mathclose\aftergroup{\aftergroup}\right}

\usepackage[export]{adjustbox}

\makeatletter
\@namedef{ver@everyshi.sty}{}
\makeatother
\usepackage{tikz}
\usepackage{pgfplots}
\usetikzlibrary{spy,calc}

\pgfplotsset{compat=1.16}

\newif\ifblackandwhitecycle
\gdef\patternnumber{0}

\pgfkeys{/tikz/.cd,
    zoombox paths/.style={
        draw=orange,
        very thick
    },
    black and white/.is choice,
    black and white/.default=static,
    black and white/static/.style={ 
        draw=white,   
        zoombox paths/.append style={
            draw=white,
            postaction={
                draw=black,
                loosely dashed
            }
        }
    },
    black and white/static/.code={
        \gdef\patternnumber{1}
    },
    black and white/cycle/.code={
        \blackandwhitecycletrue
        \gdef\patternnumber{1}
    },
    black and white pattern/.is choice,
    black and white pattern/0/.style={},
    black and white pattern/1/.style={    
            draw=white,
            postaction={
                draw=black,
                dash pattern=on 2pt off 2pt
            }
    },
    black and white pattern/2/.style={    
            draw=white,
            postaction={
                draw=black,
                dash pattern=on 4pt off 4pt
            }
    },
    black and white pattern/3/.style={    
            draw=white,
            postaction={
                draw=black,
                dash pattern=on 4pt off 4pt on 1pt off 4pt
            }
    },
    black and white pattern/4/.style={    
            draw=white,
            postaction={
                draw=black,
                dash pattern=on 4pt off 2pt on 2 pt off 2pt on 2 pt off 2pt
            }
    },
    zoomboxarray inner gap/.initial=3pt,
    zoomboxarray columns/.initial=1,
    zoomboxarray rows/.initial=2,
    subfigurename/.initial={},
    figurename/.initial={zoombox},
    zoomboxarray/.style={
        execute at begin picture={
            \begin{scope}[
                spy using outlines={%
                    zoombox paths,
                    width=\imagewidth / \pgfkeysvalueof{/tikz/zoomboxarray columns} - (\pgfkeysvalueof{/tikz/zoomboxarray columns} - 1) / \pgfkeysvalueof{/tikz/zoomboxarray columns} * \pgfkeysvalueof{/tikz/zoomboxarray inner gap} -\pgflinewidth,
                    height=\imageheight / \pgfkeysvalueof{/tikz/zoomboxarray rows} - (\pgfkeysvalueof{/tikz/zoomboxarray rows} - 1) / \pgfkeysvalueof{/tikz/zoomboxarray rows} * \pgfkeysvalueof{/tikz/zoomboxarray inner gap}-\pgflinewidth,
                    magnification=3,
                    every spy on node/.style={
                        zoombox paths
                    },
                    every spy in node/.style={
                        zoombox paths
                    }
                }
            ]
        },
        execute at end picture={
            \end{scope}
            \node at (image.north) [anchor=north,inner sep=0pt] {\phantomimage};
            \node at (zoomboxes container.north) [anchor=north,inner sep=0pt] {\phantomimage};
     \gdef\patternnumber{0}
        },
        spymargin/.initial=0.5em,
        zoomboxes xshift/.initial=1,
        zoomboxes right/.code=\pgfkeys{/tikz/zoomboxes xshift=1},
        zoomboxes left/.code=\pgfkeys{/tikz/zoomboxes xshift=-1},
        zoomboxes yshift/.initial=0,
        zoomboxes above/.code={
            \pgfkeys{/tikz/zoomboxes yshift=1},
            \pgfkeys{/tikz/zoomboxes xshift=0}
        },
        zoomboxes below/.code={
            \pgfkeys{/tikz/zoomboxes yshift=-1},
            \pgfkeys{/tikz/zoomboxes xshift=0}
        },
        caption margin/.initial=0ex,
    },
    adjust caption spacing/.code={},
    image container/.style={
        inner sep=0pt,
        at=(image.north),
        anchor=north,
        adjust caption spacing
    },
    zoomboxes container/.style={
        inner sep=0pt,
        at=(image.north),
        anchor=north,
        name=zoomboxes container,
        xshift=\pgfkeysvalueof{/tikz/zoomboxes xshift}*(\imagewidth+\pgfkeysvalueof{/tikz/spymargin}),
        yshift=\pgfkeysvalueof{/tikz/zoomboxes yshift}*(\imageheight+\pgfkeysvalueof{/tikz/spymargin}+\pgfkeysvalueof{/tikz/caption margin}),
        adjust caption spacing
    },
    calculate dimensions/.code={
        \pgfpointdiff{\pgfpointanchor{image}{south west} }{\pgfpointanchor{image}{north east} }
        \pgfgetlastxy{\imagewidth}{\imageheight}
        \global\let\imagewidth=\imagewidth
        \global\let\imageheight=\imageheight
        \gdef\columncount{1}
        \gdef\rowcount{1}
        
    },
    image node/.style={
        inner sep=0pt,
        name=image,
        anchor=south west,
        append after command={
            [calculate dimensions]
            node [image container,subfigurename=\pgfkeysvalueof{/tikz/figurename}-image] {\phantomimage}
            node [zoomboxes container,subfigurename=\pgfkeysvalueof{/tikz/figurename}-zoom] {\phantomimage}
        }
    },
    color code/.style={
        zoombox paths/.append style={draw=#1}
    },
    connect zoomboxes/.style={
    spy connection path={\draw[draw=none,zoombox paths] (tikzspyonnode) -- (tikzspyinnode);}
    },
    help grid code/.code={
        \begin{scope}[
                x={(image.south east)},
                y={(image.north west)},
                font=\footnotesize,
                help lines,
                overlay
            ]
            \foreach \x in {0,1,...,9} { 
                \draw(\x/10,0) -- (\x/10,1);
                \node [anchor=north] at (\x/10,0) {0.\x};
            }
            \foreach \y in {0,1,...,9} {
                \draw(0,\y/10) -- (1,\y/10);                        \node [anchor=east] at (0,\y/10) {0.\y};
            }
        \end{scope}    
    },
    help grid/.style={
        append after command={
            [help grid code]
        }
    },
}

\newcommand\phantomimage{%
    \phantom{%
        \rule{\imagewidth}{\imageheight}%
    }%
}
\newcommand\zoombox[2][]{
    \begin{scope}[zoombox paths]
        \pgfmathsetmacro\xpos{
            (\columncount-1)*(\imagewidth / \pgfkeysvalueof{/tikz/zoomboxarray columns} + \pgfkeysvalueof{/tikz/zoomboxarray inner gap} / \pgfkeysvalueof{/tikz/zoomboxarray columns} )
        }
        \pgfmathsetmacro\ypos{
            (\rowcount-1)*( \imageheight / \pgfkeysvalueof{/tikz/zoomboxarray rows} + \pgfkeysvalueof{/tikz/zoomboxarray inner gap} / \pgfkeysvalueof{/tikz/zoomboxarray rows} )
        }
        \edef\dospy{\noexpand\spy [
            #1,
            zoombox paths/.append style={
                black and white pattern=\patternnumber
            },
            every spy on node/.append style={#1},
            x=\imagewidth,
            y=\imageheight
        ] on (#2) in node [anchor=north west] at ($(zoomboxes container.north west)+(\xpos pt,-\ypos pt)$);}
        \dospy
        \pgfmathtruncatemacro\pgfmathresult{ifthenelse(\columncount==\pgfkeysvalueof{/tikz/zoomboxarray columns},\rowcount+1,\rowcount)}
        \global\let\rowcount=\pgfmathresult
        \pgfmathtruncatemacro\pgfmathresult{ifthenelse(\columncount==\pgfkeysvalueof{/tikz/zoomboxarray columns},1,\columncount+1)}
        \global\let\columncount=\pgfmathresult
        \ifblackandwhitecycle
            \pgfmathtruncatemacro{\newpatternnumber}{\patternnumber+1}
            \global\edef\patternnumber{\newpatternnumber}
        \fi
    \end{scope}
}

\usepackage{listings}

\definecolor{codeblue}{rgb}{0,0,1}
\definecolor{codegreen}{rgb}{0,0.6,0}
\definecolor{codegray}{rgb}{0.5,0.5,0.5}
\definecolor{codepurple}{rgb}{0.58,0,0.82}

\lstdefinelanguage{HLSL}{
  morekeywords={float, float2, float3, float4, uint, uint2, uint3, uint4, 
                int, int2, int3, int4, void, bool, matrix, vector, 
                static, const, struct, return, if, else, for, while, do},
  morekeywords=[2]{WaveGetLaneIndex, WaveReadLaneAt, cross, dot, length, max, min, asfloat, asuint}, % Intrinsics
  sensitive=true,
  morecomment=[l]{//},
  morecomment=[s]{/*}{*/},
  morestring=[b]",
  morestring=[b]'
}

\lstdefinestyle{iccv_code}{
    language=HLSL,
    basicstyle=\ttfamily\footnotesize, % Matches fontsize=\footnotesize
    commentstyle=\color{codegreen}\itshape,
    keywordstyle=\color{codeblue}\bfseries,
    keywordstyle=[2]\color{codepurple}, % Intrinsics in purple
    numberstyle=\tiny\color{codegray},
    breaklines=true,
    frame=tb,              % "frame=lines" equivalent (top and bottom)
    framerule=1pt,         % Matches framerule=1pt
    rulecolor=\color{black},
    captionpos=t,          % Puts the title/label at the top
    tabsize=4,
    showstringspaces=false
}

\definecolor{iccvblue}{rgb}{0.21,0.49,0.74}
\usepackage[pagebackref,breaklinks,colorlinks,allcolors=iccvblue]{hyperref}

\def\paperID{14190}
\def\confName{CVPR}
\def\confYear{2026}

\title{Radiance Meshes for Volumetric Reconstruction}
\newcommand{\institutes}[0]{$^1$University of California, San Diego \quad $^2$Google \quad $^3$Runway ML}

\newlength{\maxauthwidth}

\newcommand{\authorbox}[2]{\makebox[\maxauthwidth][c]{#1\,$^#2$}}
\newcommand{\aemail}[1]{\makebox[\maxauthwidth][c]{{\tt\small #1}}}

\author{
    \authorbox{Alexander Mai}{1}\\
    \aemail{amai@ucsd.edu}
    \and
    \authorbox{Trevor Hedstrom}{1}\\
    \aemail{tjhedstr@ucsd.edu}
    \and
    \authorbox{George Kopanas}{{2,3}}\\
    \aemail{gkopanas@google.com}
    \and 
    \authorbox{Janne Kontkanen}{2}\\
    \aemail{jkontkanen@google.com}
    \and
    \authorbox{Falko Kuester}{1}\\
    \aemail{fkuester@ucsd.edu}
    \and
    \authorbox{Jonathan T. Barron}{2}\\
    \aemail{barron@google.com}
}

\begin{document}

\twocolumn[{
\renewcommand\twocolumn[1][]{#1}
\maketitle
\begin{center}
    \centering
    \includegraphics[width=\linewidth]{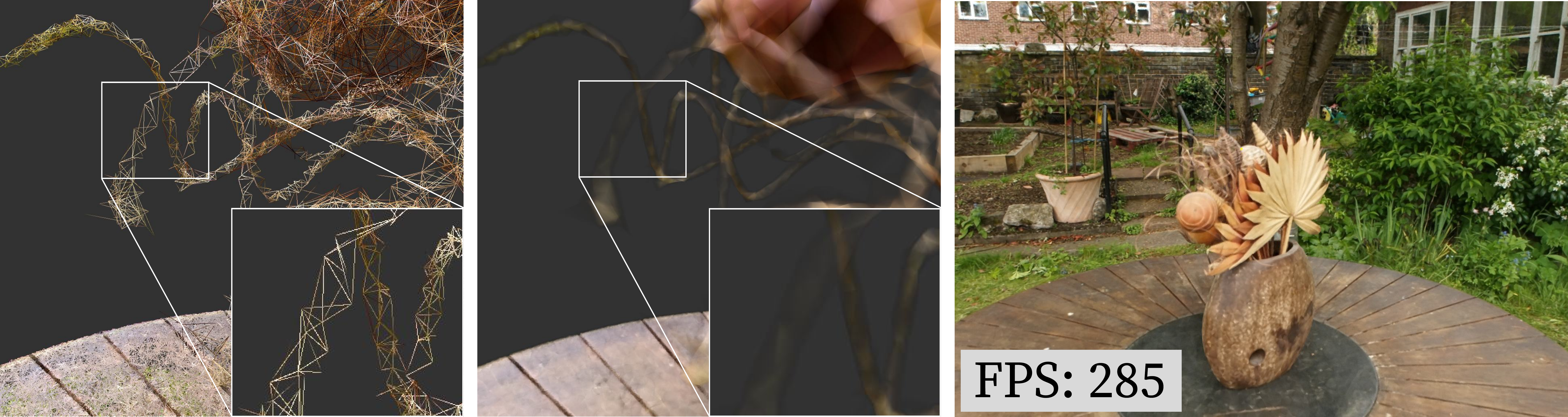}

\captionof{figure}{
A radiance mesh parameterizes a radiance field using the tetrahedra produced by a Delaunay triangulation of a set
of points, where each tetrahedron has a constant density and a linearly-varying color. This representation allows us to use traditional mesh processing (left) while still enabling semi-transparent rendering (middle), which allows our model to be rendered using a hardware triangle rasterizer very quickly on a wide variety of platforms (right). See the supplement for an interactive web demo.
}
\label{fig:frontpage}
\end{center}
}]

% Another teaser mock from another paper:
% \twocolumn[{%
% \maketitle
% \thispagestyle{empty}
% \begin{center}
% \centering
%     \includegraphics[width=\linewidth]{figures/images/teaser_mock.png}
% \captionof{figure}{\textbf{Scalable 3D reconstruction and rendering.}
% }%
% \label{fig:teaser-fig}%
% \end{center}%
% }]

\begin{abstract}
We introduce radiance meshes, a technique for representing radiance fields with constant density tetrahedral cells produced with a Delaunay tetrahedralization.
Unlike a Voronoi diagram, a Delaunay tetrahedralization yields simple triangles that are natively supported by existing hardware. As such, our model is able to perform exact and fast volume rendering using both rasterization and ray-tracing. We introduce a new rasterization method that achieves faster rendering speeds than all prior radiance field representations (assuming an equivalent number of primitives and resolution) across a variety of platforms.
Optimizing the positions of Delaunay vertices introduces topological discontinuities (edge flips). To solve this, we use a Zip-NeRF-style backbone which allows us to express a smoothly varying field even when the topology changes.
Our rendering method exactly evaluates the volume rendering equation and enables high quality, real-time view synthesis on standard consumer hardware. Our tetrahedral meshes also lend themselves to a variety of exciting applications including fisheye lens distortion, physics-based simulation, editing, and mesh extraction.

\end{abstract}

\section{Introduction}
\label{sec:intro}
Radiance fields have become the standard paradigm for 3D reconstruction since NeRF~\cite{mildenhall2020nerf}. 
It pioneered the use of differentiable volume rendering for directly optimizing a radiance field to explain a set of captured images. 
%established the practice of reconstructing scenes using differential volume rendering and gradient descent on the photometric difference between captured and rendered images.
%Since NeRF, the scientific community has explored a variety of different volumetric  representations~\cite{tewari2022advances}, each with different strengths and weaknesses.
A fundamental tradeoff persists in radiance field model design: representations that are fast to render are difficult to optimize, such as opaque triangle meshes. Conversely, representations with well-behaved optimization dynamics, like MLPs, tend to be slow to render.

A particularly effective point on this trade-off continuum is 3D Gaussian Splatting (3DGS)~\cite{kerbl20233d}. It is fast to render and relative easy to make backwards-compatible with the existing graphics ecosystem. But 3DGS has a variety of limitations: 1) The volume rendering approximations that allow the high frame-rates introduce popping artifacts. 2) Splatting is challenging with complicated camera models like ultra-wide lenses. 3) Splats can be difficult to edit. 4) Though 3DGS is faster than most radiance field models, it is significantly slower than conventional mesh-based real-time rendering techniques.
%most prominently that the approximations used introduce popping artifacts. VR is an important application, and not only are these artifacts more noticeable there, but the ultra wide lenses require sophisticated engineering efforts to splat against correctly.
% As such, papers, including this one, have continued to explore different representations to overcome the limitations of 3DGS.

%Recently, Radiant Foam~\cite{govindarajan2025radiant} introduced an approach that partitions space according to an optimizable 3D Voronoi diagram. While the approach strikes a good balance between speed and quality, it requires rendering arbitrarily complex convex \emph{polyhedra} (the Voronoi cells), which is non trivial using native GPU rasterization and instead requires the use of a slower ray tracing algorithm. While Radiant Foam enables compelling capabilities (i.e wide angle lenses), it still under performs state-of-the-art radiance field models in both quality and speed.
The recently introduced Radiant Foam~\cite{govindarajan2025radiant} partitions space using an optimizable 3D Voronoi diagram. While this approach strikes a good balance between speed and quality, it relies on ray tracing arbitrarily complex convex \emph{polyhedra} (the Voronoi cells). While GPU-based rasterization of these cells is theoretically possible, it would be prohibitively costly: Voronoi cells generally possess an average of 15.5 faces~\cite{meijering1953interface}. Rasterizing a single cell would require tessellating these faces into dozens of triangles. Even accounting for the fact that the dual Delaunay graph contains multiple tetrahedra per Voronoi site, the total triangle count required to rasterize the Voronoi diagram is nearly double that of the Delaunay representation. In addition, to find the transparency of each cell, we need to find the distance from the front face to the back face. However finding the distance to the back face of the polyhedra requires iterating over all of it's neighbors within the fragment shader, which is very expensive.

We introduce a method that is able to perform high quality view synthesis using tetrahedral meshes - specifically, a Delaunay tetrahedralization. The Delaunay tetrahedralization partitions space into tetrahedral cells, to which we assign a constant density and linear color (radiance) to create what we call a \emph{radiance mesh}.
The use of Delaunay tetrahedralization enables unconstrained optimization of the vertex locations, as we can simply recompute the triangulation after the points have moved.
RadFoam had considered the use of a Delaunay tetrahedralization, as it is the dual of the Voronoi diagram. However, they chose not to pursue it because of the discontinuous mapping from vertex positions to mesh topology. These present difficulties during optimization, which we overcome.

The resulting radiance mesh has a powerful property, which is that it can be sorted front to back very efficiently~\cite{edelsbrunner1989acyclicity, max1990area, karasick1997visualization}, using a radix sort over the power of the circumsphere. This, combined with the fixed number of faces per cell, enables fast rasterization, even faster than Gaussian splatting, but without the popping. Since the tetrahedral mesh produced is comprised entirely of triangles, it is possible to utilize the hardware triangle rasterizer to accelerate rendering, and to utilize the hardware triangle intersector to accelerate ray tracing.
For this purpose, we introduce a novel hardware rasterization method that interpolates the exit intersection and uses mesh shaders to collaboratively load data onto vertices.
We achieve rendering speeds 32\% faster than the original 3DGS implementation at 1440p resolutions and higher quality than RadFoam. For ray tracing, we achieve a 17\% speed increase compared to RadFoam on MipNeRF360 indoor scenes. Crucially, unlike 3DGS, our method utilizes exact visibility. This guarantees that radiance meshes are free from the temporal popping and photometric inconsistencies caused by the sorting errors inherent to splatting.

Because radiance meshes consist of semi-transparent triangles, they are natively compatible with the real-time graphics ecosystem. 
For example, we can support physics-based simulations using Extended Position Based Dynamics~\cite{macklin2016xpbd}, which we demonstrate as an interactive application.
Additionally, we can extract a watertight opaque surface triangle mesh by thresholding our primitives according to their contribution to the rendered views.
We provide reference implementations for both the rasterizer on desktop and web, as well as the ray tracer. 

\section{Related Work}
\label{sec:related_work}

\myparagraph{Neural Volume Rendering}.
Neural Radiance Fields use numerical quadrature to integrate emission-only volume rendering parameterized by a neural network~\cite{mildenhall2020nerf}. Many improvements to NeRF have since been proposed, such as different filtering approaches~\cite{barron2021mip, barron2022mipnerf360, barron2023zipnerf} and different field representations ~\cite{chan2022efficient, fridovich2022plenoxels, mueller2022instant, lombardi2021mixture, xu2022point}. Adaptive Voronoi NeRFs proposed partitioning a scene into multiple radiance fields using a Voronoi diagram~\cite{elsner2023adaptive}, and
Radiant Foam~\cite{govindarajan2025radiant} partition a scene into constant density Voronoi cells to enable fast ray-tracing.

\myparagraph{Particle-Based Radiance Fields}.
Early works~\cite{lassner2021pulsar, aliev2020neural, wiles2020synsin} explored particle based representations as an alternative to fields. 3DGS~\cite{kerbl20233d} popularized them by using Gaussian primitives. Many other primitives have since been explored \cite{liu2025deformable, hamdi2024ges, qu2024disc, li20243d, held20253d, huang2025deformable, zhang2024quadratic, Chen_2024, held2025triangle, huang20242d}, largely based on the same splatting-based alpha compositing technique as 3DGS.
Most splatting-based methods make different approximations to volume rendering that result in instabilities and ``popping''. StopThePop~\cite{radl2024stopthepop} and 3DGRT~\cite{3dgrt2024} use per-ray-sorting to improve popping. SVRaster~\cite{sun2025sparse} is able to perform high speed rasterization of voxel primitives using few quadrature samples within each primitive. EVER~\cite{mai2024ever} is the first primitive-based representation able to perform fast and exact integration via ray-tracing.

\myparagraph{Mesh-Based View synthesis}.
Because meshes are the dominant representation used in computer graphics, many have explored how to best use meshes to accelerate radiance field rendering.
Popular approaches include optimizing a field that has a dual interpretation and can be used both as a radiance field and a surface~\cite{yariv2021volume,yariv2023bakedsdf, wang2021neus}.
Another way to obtain mesh-like properties while retaining the image quality of 3DGS is to attach Gaussians to a mesh~\cite{waczynska2024games,lin2024direct,qian2024gaussianavatars,shao2024splattingavatar,choi2024meshgs,zheng2025gaustar,gao2025mani,wang2024gaussurf,gao2024real}, convert Gaussians to a mesh using a Delaunay triangulation~\cite{guedon2025milo, yu2024gaussian}, or binarize during the optimization a volumetric field such that it can be converted to a mesh~\cite{reiser2024binary}. 
Meshes can also be directly optimized via gradient descent~\cite{chen2023mobilenerf, kato2018neural, li2018differentiable}, though these approaches are limited due to edge discontinuities, which cause optimization issues. 
% like in NVDiffRast~\cite{laine2020modular} or NVDiffRec~\cite{munkberg2022extracting}, using SDFs.

\myparagraph{Tetrahedralizations}.
Tetrahedra has been widely studied as a volumetric representation in graphics, specifically for solids. To accelerate rendering, power sorting was first introduced by \citet{karasick1997visualization}. Even further acceleration were introduced by~\cite{museth2004tetsplat, georgii2006generic} and ~\citet{tricard2024interval} utilized geometry shaders for further acceleration. Tetrahedra have also been used for view synthesis in ~\cite{von2025linprim, guo2024tetsphere}.

% =====================================================================
%  Differentiable Tetrahedral Volume Rendering (abstract overview)
% =====================================================================
\section{Preliminaries}

\myparagraph{Neural Radiance Fields} 
A radiance field is an emission-only volumetric field consisting of a density field $\sigma(\mathbf{x})$ and a color field $\colorc(\mathbf{x}, \mathbf{d})$, where color may depend on viewing direction $\mathbf{d}$.

Each pixel’s color is rendered by setting up a ray $\ray(t) = \mathbf{o} + t\mathbf{d}$ with origin $\mathbf{o}$ and direction $\mathbf{d}$ through the pixel, and ray marching using standard volume rendering integral (also known as the radiative transfer 
%Given a ray $\ray(t) = \mathbf{o} + t\mathbf{d}$ with origin $\mathbf{o}$ and direction $\mathbf{d}$, the pixel-color of the ray is rendered using standard volume rendering integral (also known as the radiative transfer
equation~\cite{chandrasekhar1960radiative}):
\begin{equation} \label{eqn:volumerendering}
    \int_0^\infty \colorc \lft(\ray(t), \mathbf{d} \rgt) \sigma\lft(\ray(t)\rgt) \exp\lft(-\int_0^t \sigma(\ray(s))\,ds\rgt)\,dt \, .
\end{equation}
% The parameterization of the density and color fields can vary from small MLPs~\cite{mildenhall2020nerf} to large hierarchies of hashes and grids~\cite{mueller2022instant}. 

\myparagraph{Delaunay Triangulation} \label{sec:delaunay}% \& Power Sorting}
% An N-d Delaunay triangulation~\cite{delaunay1934sphere} is a point set triangulation, defined as a triangulation of the point set $P$ such that all circum-hypersphere corresponding to the $d$-simplices are empty, that is, they don't contain points in $P$.
% The circum-hypersphere of the $d$-simplex is the hypersphere uniquely defined as having the surface intersect the vertices of the $d$-simplex. 
Given a set of points, the Delaunay triangulation~\cite{delaunay1934sphere} is the triangulation defined by the empty sphere property: that the circumscribed sphere (circumsphere) corresponding to every tetrahedron contains no other vertices. 
%The circumsphere is uniquely defined as the smallest sphere containing the tetrahedron. \jk{no need to explain more. this is pretty common terminology}
A Delaunay tetrahedralization can be sorted from front to back~\cite{max1990area, edelsbrunner1989acyclicity} with respect to a camera origin $\bf{o}$ according to the power of each circumsphere with respect to the camera origin:
\begin{equation}
P(T) = \|\circumcenter(T) - \mathbf o\|_2^2 - \circumradius(T)^2\,,
\label{eq:power_sort}
\end{equation}
where $\circumradius(T)$ and $\circumcenter(T)$ are the radius and center, respectively, of the circumsphere of $T$.
We offer a visual intuition of why this works using the radical axis in the supplement while the full proof can be found in~\cite{karasick1997visualization}. This sort order is particularly useful because it is with respect to the ray origin, which means it is valid for fisheye lens and other such distorted projections.

\myparagraph{Zip-NeRF Downweighting}
%Instant-NGP~\cite{mueller2022instant} represents a 3D function $G_\theta$ using multiple feature grids at different resolutions, ranging from very coarse to extremely fine. For the larger grids, it uses a hash function to compress these high-resolution grids into a much smaller, fixed-size table of features, relying a neural network to learn to resolve collisions.
To mitigate aliasing, Zip-NeRF~\cite{barron2023zipnerf} introduced a querying scheme where the output of the instant-NGP is convolved with a gaussian. For a given query with $\sigma$ denoting the filter variance, it multiplies each level $\ell$ with a grid resolution of $n_\ell$ by the following factor:
\begin{equation}
\phi(\sigma) = \text{erf}\lft(1/\sqrt{8\sigma^2n_\ell^2}\rgt)\,,
\end{equation}
Together with a regularizer that pulls all features towards $0$, this reduces the impact of higher resolution features for larger sized queries.

\section{Method}
\label{sec:model}
Our model takes as input a set of posed images and sparse initial point cloud. 
During optimization, we represent our scene using a set of 3D points $\{ \mathbf{v}_i \}$, initialized from the COLMAP SfM reconstruction~\cite{schoenberger2016sfm}, and an Instant-NGP datastructure $G_\theta$.
Every 10 optimization iterations, the Delaunay triangulation of points $\{ \mathbf{v}_i \}$ is computed to yield a set of non-overlapping tetrahedra $\{T_k\}$.
Because $\{ \mathbf{v}_i \}$ move during optimization, tetrahedra may appear and disappear. We therefore use the Instant-NGP data structure to parameterize the non-spatial properties of each tetrahedron. The output of our model is a set of tetrahedra $\{T_k\}$, each with a constant density $\sigma_k$, a base color given by spherical harmonics $\colorc_k^0$, and a linear color gradient $\cgradient_k$, all of which are produced by querying $G_\theta$ with each $T_k$ (see Section~\ref{sec:query}).
% To explain how we construct a differentiable and smooth volumetric field using a Delaunay tetrahedralization, before explaining how we can render it at a high frame-rate.

% \subsection{Representation \& Rendering}

% During rendering, we assign each tetrahedral cell a constant density.
% However, while an approach like EVER~\cite{mai2024ever} can utilize color blending between overlapping primitives to produce smooth color changes, our primitives do not overlap, so we need a different method to allow for smooth shading. 
% We chose to linearily vary the color across each tetrahedron. Each tetrahedron is given a vector $\cgradient_k\in R^3$, which is a monochromatic vector that describes how all of the colors change across the primitive.
Given our parameterization, the color at any interior point \(\mathbf p\in T_k\) is:
\begin{equation}
  \mathbf c_k(\mathbf p) = \colorc_k^0+\cgradient_k \cdot 
    \bigl(\mathbf p-\circumcenter_k\bigr)\,,
  \label{eq:linear-colour}
\end{equation}
where $\circumcenter_k$ is either the circumcenter or centroid of $T_k$.
% Let \(\Delta t=t_{\mathrm{out}}-t^{\mathrm{in}}_k\), obtained using the Cyrus-Beck line clipping algorithm, and define the entry/exit colors

\subsection{Rendering}
Before rendering, we sort $\{T_k\}$ utilizing power sorting (Equation ~\ref{eq:power_sort}).
For any given ray, we can now iterate over each primitive and compute the emission only volume rendering equation across the interval bounded by the primitive.
Let $t^{\mathrm{in}}_k$ and $t^{\mathrm{out}}_k$ be the distances where the ray $\mathbf{r}(t)$ enters and exits the $k$'th tetrahedron (computed using the Cyrus-Beck algorithm~\cite{cyrus1978}).
The color contribution of tetrahedron $k$ to the ray, premultiplied by alpha, has the form:
% Given density \(\sigma_k\) is constant along the segment, the volume-rendered color contribution is of the $k$'th tetrahedron to the ray's pixel is:
% \begin{equation}
%   \mathbf{\Delta C}_k
%   =\int_{0}^{\Delta t}\!
%      e^{-\sigma_k t}\,\mathbf C_k\!\bigl(\mathbf r(t^{\mathrm{in}}_k+t)\bigr)\,
%      \sigma_k\, dt \,.
% \end{equation}
\begin{equation}
  \mathbf{\Delta C}_k
  =\int_{t^{\mathrm{in}}_k}^{t^{\mathrm{out}}_k}\!
     e^{-\sigma_k (t - t^{\mathrm{in}}_k)}\,\mathbf c_k\!\bigl(\mathbf r(t)\bigr)\,
     \sigma_k\, dt \,.
\end{equation}
This integral has a closed form solution that depends only on the constant density and the color of the tetrahedron at its entry and exit points. This solution for the color premultiplied by the alpha is as follows:
\begin{gather}
  \mathbf{\Delta C}_k = 
  \lft( 1 - \frac{\alpha_k}{d_k} \rgt) \mathbf{c}_k^\mathrm{in} + 
  \lft( \frac{\alpha_k}{d_k} - e^{-d_k}  \rgt) \mathbf{c}_k^\mathrm{out}\,\label{eq:segment-integral}, \\
  \begin{aligned}
  d_k &= \lft(t^{\mathrm{out}}_k-t^{\mathrm{in}}_k\rgt) \sigma_k \quad&\quad \alpha_k &= 1-e^{-d_k} \label{eq:segment-alpha} \\ \mathbf{c}_k^\mathrm{in} &= \mathbf{c}_k \lft(\ray\lft(t^{\mathrm{in}}_k\rgt)\rgt) \quad&\quad \mathbf{c}_k^\mathrm{out} &= \mathbf{c}_k \lft(\ray\lft(t^{\mathrm{out}}_k\rgt)\rgt)
  \end{aligned}
\end{gather}
To render the radiance mesh, we rasterize the tetrahedra front-to-back, using Eq. \ref{eq:segment-integral} and Eq. \ref{eq:segment-alpha} to compute the color and opacity of each tetrahedron, respectively. The final pixel color is found by blending the colors of the pre-sorted tetrahedra along the ray as follows:
\begin{equation}
    \mathbf{C} = \sum_k w_{k} \mathbf{\Delta C}_k\,, \quad\quad w_{k} = \prod^{k-1}_{l=1} \lft( 1-\alpha_\ell \rgt) \,,
\end{equation}
which we efficiently compute using hardware rasterization. For each pixel, the $\mathbf{\Delta C}_k$ and $\alpha_k$ values are computed once per tetrahedra within the fragment shader, and the product of opacities $w_k$ is efficiently computed using hardware blending operations.

\subsection{Querying Tetrahedron Properties}
\label{sec:query}

As vertices move during optimization, tetrahedra might appear and disappear after recomputing the Delaunay Tetrahedralization. Attaching density and appearance features to vertices works poorly, as we demonstrate in our ablation study in Tab.~\ref{tab:ablations}.
We therefore optimize a field, parameterized by an Instant-NGP datastructure~\cite{mueller2022instant} and the mip-NeRF 360 contraction function~\cite{barron2022mipnerf360}, which we query with each tetrahedron to produce density and appearance features.

\begin{figure}
    \centering
    \includegraphics[width=\linewidth]{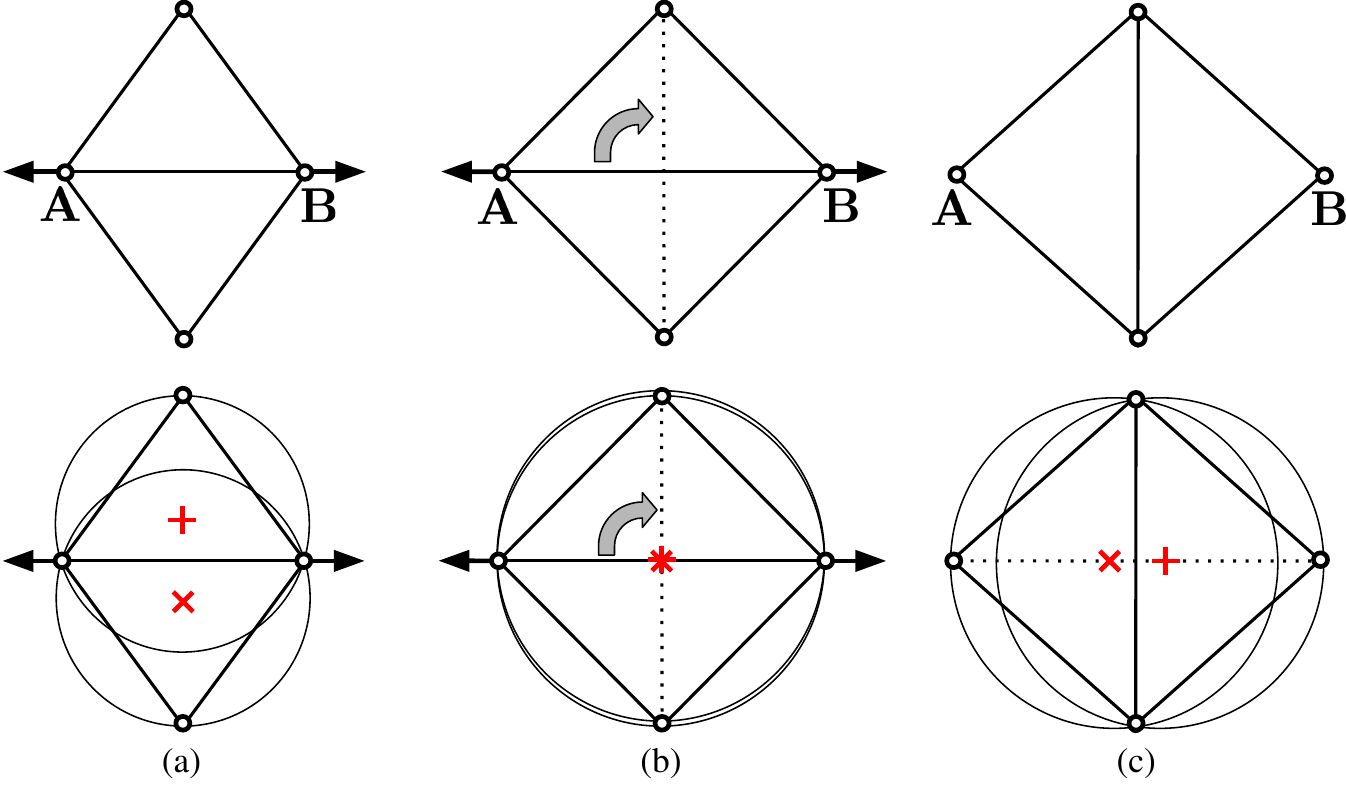}
    \caption{
    {\bf Top}: Vertices $\mathbf{A}$ and $\mathbf{B}$ are moved slightly as we go from left to right, illustrating the the problem with Delaunay Triangulation. This small modification causes a topological flip. Thus naively assigning colors to vertices and using barycentric interpolation would not give a smooth optimization objective. {\bf Bottom}:  Our circumcenter smoothing. We color each triangle based on the position of their circumcenters (\textcolor{red}{+} and \textcolor{red}{$\times$}). This works, because at the moment of the flip the centers are co-located.   
    % https://docs.google.com/drawings/d/1JSmYMPD1XZQuhUrhiVo_0WotGG99HOGBuuMIRpULWT0/edit?usp=sharing
%    Two different approaches for assigning attributes (here, color) to the same Delaunay triangulation, before (left) and after (right) an edge flip.
 %   Top: The naive approach attaching colors to vertices and using the average of the vertices for each tetrahedron means that, when topology changes, the field goes from being blue and green to blue-green. Bottom: our approach of coloring triangles according to their circumcenter locations (shown here as circles) means that, when vertices approach a point where their topology can undergo a sudden flip, both tetrahedra are guaranteed to already have the same color, and the edge-flip does not induce a color change.
%    \jb{this figure would be much easier to understand if the left and right centroids and circumspheres were indicated using different symbols from each other. The reader isn't going to understand that the one circle in the middle is actually two circles on top of each other. Using an x and a + would work great. also there's probably no value in rendering the centroids or circumcenters of all the other tets, they are just confusing and do not convey the information this figure is intended to convey}
    }
    \label{fig:delaunay_smoothing}
\end{figure}

%Because tetrahedra are volumes but Instant-NGP expects points as input, we query the NGP using an approach similar to that of Zip-NeRF~\cite{barron2023zipnerf}.
%Critically, we use the circumcenter $\circumcenter_k$ of each tetrahedron $k$ to query the Instant-NGP, which results in a smooth and continuous function during optimization in the case of edge-flipping, visualized in Fig.~\ref{fig:delaunay_smoothing}. One catch is that, when a tetrahedron becomes skinny, its vertices become nearly coplanar, which makes the system of linear equations used to find the circumcenter ill-conditioned (or nearly singular). 
%To mitigate this, we use its circumradius $\circumradius_{k}$ in Zip-NeRF's downweighting function $\phi$ to effectively prefilter the Instant NGP. As the matrix nears singular, the circumradius grows, increasing the smoothing effect.
Because tetrahedra are volumes but Instant-NGP expects points as input, we query the NGP using an approach similar to that of Zip-NeRF~\cite{barron2023zipnerf}.
We evaluated two strategies as the center $\circumcenter_k$ of a tetrahedra: the circumcenter and the centroid. The circumcenter has the advantage of making the method continuous. This is because,the circumcenters are guaranteed to be the same for tetrahedra involved in edge-flipping as visualized in Fig.~\ref{fig:delaunay_smoothing}, resulting in a continuous function over the Delaunay Tetrahedralization.
However, the circumcenter has poor numerical conditioning, as it can move very quickly when a tetrahedron is skinny.
The circumradius $\circumradius_k$ is used to prefilter the NGP to address skinny tetrahedra. As a tetrahedron becomes more co-planar, the system of linear equations used to find the circumcenter becomes poorly conditioned (nearly singular), and the circumradius becomes larger. By filtering the NGP using this radius, using downweighing as in Zip-NeRF~\cite{barron2023zipnerf}, we attempt to mitigate this issue.
Our second strategy, the centroid, has the disadvantage of not being continuous, but has much better numerical conditioning. We attempt to counteract this lack of continuity with the same filtering strategy.  We compare these two methods in the ablation and show that the centroid has better performance.

%Because tetrahedra are volumes but Instant-NGP expects points as input, we query the NGP using an approach similar to that of Zip-NeRF~\cite{barron2023zipnerf}. We use a combination of the circumcenter $\circumcenter_k$ and its circumradius $\circumradius_{k}$ to recover a smooth and continuous function during the optimization. The circumcenter is guaranteed to be the same for tets during edge-flipping as visualized in Fig.~\ref{fig:delaunay_smoothing}, guaranteing a continuous function over Delaunay Tetrahedrazilations. The circumradius addresses the pathological case of skinny tetrahedrons. Co-planar vertices make the system of linear equations used to find the circumcenter ill-conditioned (or nearly singular). The circumradius prefilters the NGP and as the matrix nears singular, the circumradius grows, increasing the smoothing effect.

% Our second strategy, the centroid, has the disadvantage of not being continuous, but has much better numerical conditioning. We attempt to counteract this lack of continuity with the same filtering.

Each individual attribute is produced by the Instant NGP $G_\theta$ with downweighting function $\phi$ as follows:
\begin{align}
    \mathbf{b}_k &\leftarrow \phi(\circumradius_{k}) \circ G_\theta\lft(\circumcenter_k\rgt)\,, \\
    \sigma_k &\leftarrow \exp(h_\sigma(\mathbf{b}_k))\,, 
    \\
    \colorc_k^0 &\leftarrow\operatorname{softplus}_{\beta=10}\lft(
     h_{\sh}\lft( \mathbf{b}_k \rgt) \cdot Y \lft( d_k \rgt) 
     \rgt)\,,
     \\
    \cgradient_k &\leftarrow 
    \frac{
    \min{\colorc}_k
    }{
    \circumradius_{k}
    } \frac{
     h_\Delta(\mathbf{b}_k))
    } {
    \sqrt{1+\|h_\Delta(\mathbf{b}_k))\|_2^2}
    }
    \,,\label{eq:color_gradient_act} % \in \Re^3 (JB: I commented this out because the notation is weird, and also the c_k is  bold which implies it's a vector, which implies it's 3D
\end{align}
where each of $h_\sigma$, $h_\Delta$, $h_c$, and $h_{\sh}$ are the different shallow neural network heads that take the Instant-NGP bottleneck feature $\mathbf{b}$ as input, and where $Y(\mathbf{d}_k)$ are the spherical harmonics basis functions evaluated in the direction $\mathbf{d}_k$.

The purpose of the activation on the color gradient $\cgradient_k$ is 
to prevent the color from going negative within the tetrahedron. 
Since the vertices of the tetrahedron lie within a ball of radius $\circumradius_k$, 
combining Eq.~\ref{eq:color_gradient_act} into Eq.~\ref{eq:linear-colour} yields a maximum negative change of $-\min{\colorc}_k$. We need to use the minimum across channels because we reuse the color gradient across all color channels for the purpose of efficiency, instead of giving each channel it's own gradient. Using a monochrome gradient saves 30\% of memory loads in the shaders.

During optimization we follow Zip-NeRF and regularize our model using the normalized $L_2$ weight decay on Instant-NGP grid. We also use the distortion loss from mip-NeRF360~\cite{barron2022mipnerf360} . We impose this on density, rather than on the alpha compositing weights, because the original formulation is not designed for analytical integration.

\subsection{Mesh Densification}
\label{sec:densification}

Particle-based radiance field methods like 3DGS use heuristics to densify or prune particles based on their contribution to rendered images. We use only densification, done by adding points to $\{\vertex_i \}$ during optimization.
We conceptualize the densification process as choosing a tetrahedron to ``split'', then choosing a point within the selected tetrahedron as the insertion point.
Every 500 iterations, we iterate over a sample of the training dataset of size $M$ and collect two metrics: an SSIM split score and a total variance split score, select primitives to ``split'' according to these scores, then chose a point to add to $\{\vertex_i \}$ from each selected primitive, before continuing with optimization. 
 
The first and most important selection method, the SSIM split $\mathcal{S}_k$, is based on Error Based Densification~\cite{rota2024revising}.
We select all primitives with a score $\mathcal{S}_k>0.5$ for densification.
The idea is to back-project the image error associated with each pixel, specifically SSIM, onto all of the primitives that contributed to that pixel according to the proportion they contributed, or the weight. These are accumulated on the primitives across the image to obtain an error score, then we take the mean of the top two error scores across the sample images $\{\pi_n\}_{n=1}^M$:
\newcommand{\allind}{i}
\begin{gather}
    \nu_k^{\pi_n} = 
    \left[
    \sum_{\allind} \left\{
        \mathbbm{1}[w_k \alpha_k > 0]
    \right\}_{\allind}
    \right]_{\pi_n} \\
    \mathcal{S}_k = 
    \operatorname{mean}\!\lft( \!\operatorname*{top2}_{\pi_n} \left[
    \frac1{\nu_k^{\pi_n}}
    \sum_{\allind}
    \left\{
        w_k \alpha_k s
    \right\}_{\allind}
    \right]_{\pi_n} \rgt) \nonumber
\end{gather}
The brackets $\left[\{\cdot\}_\allind\right]_{\pi_n}$ indicate that everything within takes place within camera $\pi_n$ at pixel $i$. So $\left[\sum_i\right]_{\pi_n}$ sums over all pixels in image $\pi_n$, and $s$ is the SSIM error at pixel $i$ within camera $\pi_n$. 
% We have to divide by the normalization factor of $\sqrt{\nu_k}$ as a heuristic because otherwise we over densify primitives right in front of the camera. 
The top two scores are used because at least two views are needed to triangulate a primitive, and there is likely little value to densifying a primitive that cannot be triangulated.

%This first metric does the majority of the densification, but is only really capable of densifying regions where the model already has something high density. 

This first metric drives the majority of the densification decisions, but fails to detect regions that are low density to begin with.
For regions with thin structures represented by low density early in optimization, we devise a second selection strategy we call the total variance split score $\mathcal{T}_k$, with the selection of all primitives $\mathcal{T}_k>2.0$ for densification. We treat each tetrahedron as an estimator of radiance within its volume and analyze the variance of its residuals. We compute the variance of the error at each tetrahedron to measure whether the radiance is being represented well. This variance, $\sigma_k^2$, is then multiplied by the weighted sum of the  total error, as follows:
\newcommand{\resid}{\boldsymbol{\delta}}
\newcommand{\expectation}{\mathbb{E}}
\begin{gather}
    % \sigma^2_k = 
    %  \frac{
    %  \sum_{\pi_n} \left[
    %  \sum_i \left\{\alpha_kw_k\resid^2\right\}_i 
    %  \right]_{\pi_n}
    %  }{
    %  \sum_{\pi_n} \left[
    %  \sum_i \left\{\alpha_kw_k\right\}_i
    %  \right]_{\pi_n}
    %  } - \left( \frac{
    %  \sum_{\pi_n} \left[
    %  \sum_i \left\{\alpha_kw_k \resid\right\}_i
    %  \right]_{\pi_n}
    %  }{
    %  \sum_{\pi_n} \left[
    %  \sum_i \left\{\alpha_kw_k\right\}_i
    %  \right]_{\pi_n}
    %  } \right)^2 
    \sigma^2_k = 
     \expectation[\resid^2] - \expectation[\resid]^2\,, \quad 
    \expectation[\mathbf{x}] = 
    \frac{
    \sum_{\pi_n}\left[
    \sum_i \left\{\alpha_k w_k \mathbf{x} \right\}_i
    \right]_{\pi_n}
    }
    {
    \sum_{\pi_n}\left[
    \sum_i \left\{\alpha_kw_k \right\}_i
    \right]_{\pi_n}
    }  \nonumber
    \\
    \mathcal{T}_k = \sigma_k^2
    \sum_{\pi_n}\left[
    \sum_i \left\{\alpha_kw_k\right\}_i
    \right]_{\pi_n}
\end{gather}
% \newcommand{\renderfun}{\operatorname{render}}
% \begin{gather}
%     \sigma^2_k = 
%      \expectation\lft[\resid^2 \rgt] - \expectation\lft[\resid \rgt]^2\,, \quad 
%      \mathcal{T}_k = \sigma_k^2 \renderfun(|\resid|)
%     \\
%     \renderfun[\mathbf{x}] = 
%     \sum_{\pi_n}\left[
%     \sum_i \left\{\alpha_k w_k \mathbf{x} \right\}_i
%     \right]_{\pi_n} \quad 
%     \expectation[\mathbf{x}] = 
%     \frac{
%     \renderfun(\mathbf{x})
%     }
%     {
%     \renderfun(1)
%     }  \nonumber
% \end{gather}
where we use the same sum and bracket notation, and $\resid = \mathbf{C} - \mathbf{C}_{\mathrm{gt}}$ is the pixel residual.

Next, we need to select the locations of the new points within the selected tetrahedra. We note that uniform sampling within a tetrahedron is ineffective for this purpose. For example, it is common during initialization that a large tetrahedron just touches the surface that needs to be densified. In such case, a random point would almost always fall into the empty space rather than at the surface.

To address this, we select two rays corresponding to the views with highest error scores, and use their approximate intersection as the new point, falling back to a random point within the tetrahedron should the intersection lie outside of the tetrahedron.
We use the mean ray for each of the two views, computed by finding the line segment between the entrance and exit centroids, $\centroid_k^{\mathrm{in}}$ and $\centroid_k^{\mathrm{out}}$.
For a given camera $\pi_n$ and primitive $k$, those are computed as follows:
\begin{equation}
    \centroid_k^{\mathrm{in}} = \frac{
    \sum_\allind \lft\{w_k \alpha_k \mathbf{r}(t^{\mathrm{in}}_k)\rgt\}_\allind
    }{
    \sum_\allind \lft\{w_k \alpha_k\rgt\}_\allind
    },\,\hspace{3pt}  \centroid_k^{\mathrm{out}} = \frac{
    \sum_\allind \lft\{w_k \alpha_k \mathbf{r}(t^{\mathrm{out}}_k)\rgt\}_\allind
    }{
    \sum_\allind \lft\{w_k \alpha_k\rgt\}_\allind
    }
\end{equation}

We then take the two segments corresponding to the top two SSIM error cameras and find the approximate intersection. This consists of finding the midpoint between the shortest line between the two segments. Finally, this midpoint is the new vertex added to $\{ \mathbf{v}_i \}$.

\begin{figure*}[t!]
    \centering
    \includegraphics[width=0.9\linewidth]{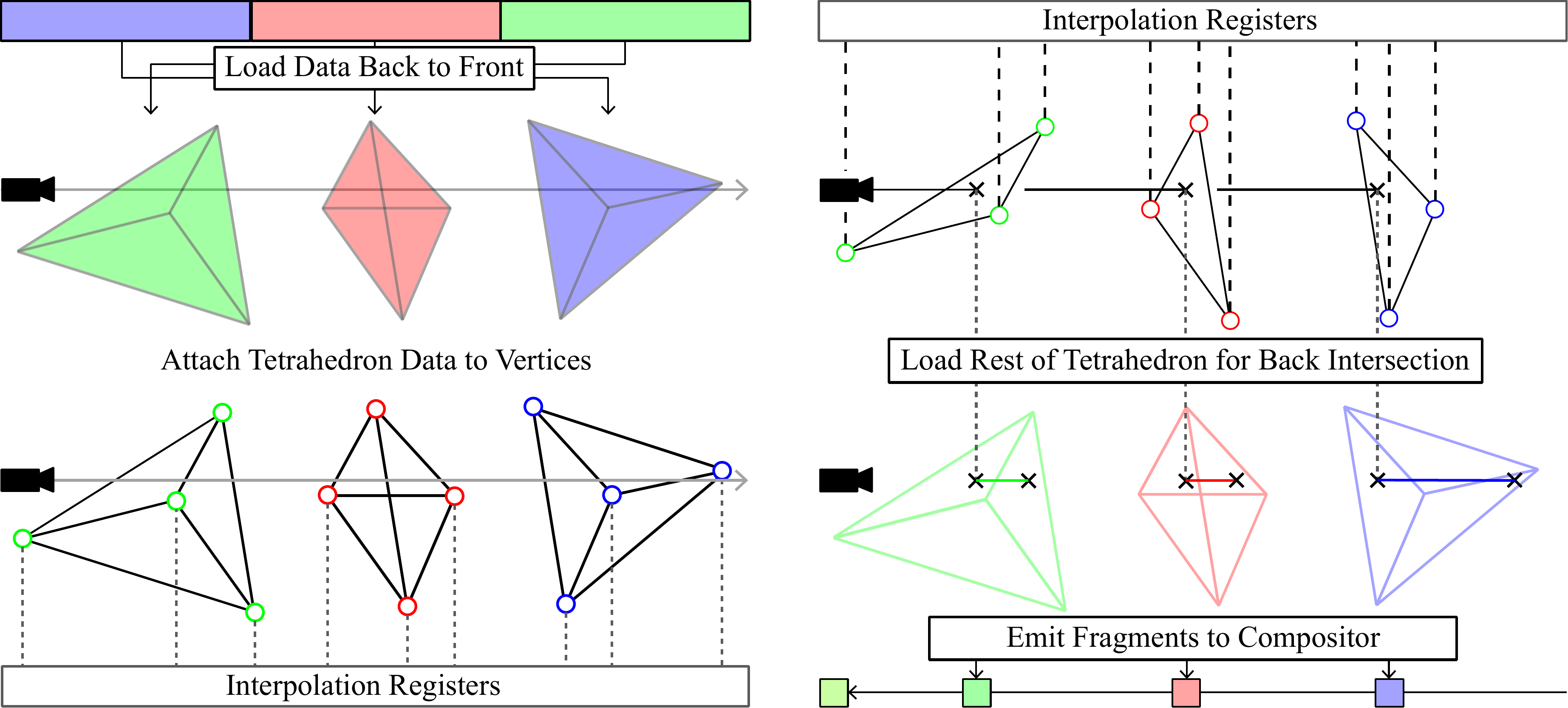}
    \caption{Radiance meshes can be efficiently rendered using the hardware rasterization on commodity GPUs. We sort all tetrahedra using the power of their circumspheres relative to the camera origin, and load them into the vertex shader from back to front. We then traverse the tets from back to front and for each, we emit 4 vertices with the tetrahedra data attached to each to the interpolation registers. This data is then read by the fragment shader, which must then read back the whole tetrahedron attached to each triangle it hit to reconstruct each back intersection. With this information, the shader can finally integrate across the interval and emit the fragment to the compositor, which composites the fragment according to the load order, so back to front.}
    \label{fig:shader_diagram}
\end{figure*}

% We supplement this processing by cloning vertices whose optimization momentum exceeds a threshold.

\subsection{High Speed Rendering}

We introduce a novel method for rasterizing a tetrahedral volume using hardware rasterization. A previous approach emitted 2-3 triangles to represent the tetrahedron~\cite{tricard2024interval}, then interpolated the transparency across each triangle. However, emitting variable amounts of geometry is slow, so instead we emit duplicate but static geometry. This includes duplicating each face twice and each vertex as many times as it is connected to a tetrahedron. We then rely on the rasterization pipeline to filter out what it needs. An overview of the approach can be see in Fig.~\ref{fig:shader_diagram}.
Our approach can be thought of as treating each tetrahedron as a distinct primitive, for which we output an individually wound tetrahedron. Then, instead of interpolating transparency, we interpolate the distance to the back faces. Details of how to interpolate this distance can be found in the Supplement.

% To take advantage of the hardware rasterizer, we need to load the faces of our tetrahedra into the vertex shader, then attach the tetrahedron data to those vertices. The hardware rasterizer will then trigger the fragment shader on all of the pixels corresponding to those triangles, linearly interpolating the vertex data to determine the appearance of the triangle. 
% If we load sorted data in the vertex shader, then the fragments outputted by the vertex shader will be composited the correct order, which means that when we go to alpha blend all of the triangle fragments, we can obtain the correct result.
% If we do so naively, each vertex is attached to multiple tetrahedra, and those tetrahedra have different attributes, hence we must duplicate each vertex once for each tetrahedron it is connected to.
% In addition, we need to know the distance to the next triangle in the tetrahedron. This means that every vertex passed to the fragment shader must contain all information about the tetrahedron it is connected to, making the process of loading data into the shaders tricky.

% The first improvement is to use
% But within the vertex shader, there are multiple possible approaches for loading the data into the duplicated vertices.
When mesh shaders are not available, we use instanced shaders, drawing an instance of a triangle strip for each tetrahedron, where the vertex shader uses the instance index to load the tetrahedron attributes and plane equations for each tetrahedron. This means that we load each tetrahedron 4 times: once for each of its vertices.
These duplicated loads can be removed by utilizing mesh shaders.
Mesh shaders replace vertex shaders, allowing an entire thread block to collaboratively generate mesh data which is passed directly to the hardware rasterizer. We leverage this feature to allow vertices to share the data they load within each tetrahedron using warp shuffle operations, so we only need to load each tetrahedron once.
%Each thread working on the 4 vertices within a tetrahedron can share data with each other, which allows us to greatly reduce the number of duplicate loads.

The differentiable PyTorch renderer is implemented as a tile based renderer in Slang.D~\cite{bangaru2023slangd} to enable tracking of derivatives, which is difficult across the vertex and fragment shaders.
The Vulkan renderer is implemented in C++ and Slang, which allows it to support mesh shaders, our fastest rendering approach. Finally, the web viewer implements instanced rendering in JavaScript and WebGL, as mesh shaders are not yet available.

\section{Results}
We evaluate radiance meshes on the nine scenes introduced in Mip-NeRF 360~\cite{barron2022mipnerf360} and four scenes from DeepBlending~\cite{hedman2018deep} and Tanks\&Temples~\cite{knapitsch2017tanks}. We compare against several recent state-of-the-art radiance field representations including 3DGS~\cite{kerbl20233d}, EVER~\cite{mai2024ever}, SVRaster~\cite{sun2025sparse}, Radiant Foam~\cite{govindarajan2025radiant}, and Triangle Splatting~\cite{held20253d}. Results are shown in Table~\ref{tab:main_results}.

\begin{figure}[h]
    \centering
    \includegraphics[width=\linewidth]{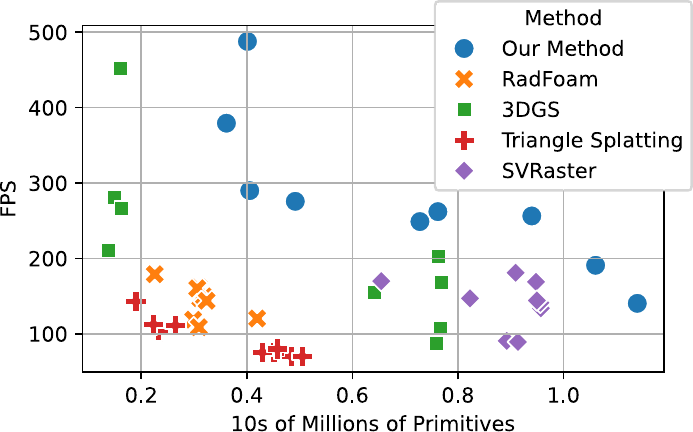}
    \caption{Rendering speed vs. primitive count. Each point represents a different scene from the Mip-NeRF 360 dataset, plotted by the number of primitives used for reconstruction ($x$-axis) against the average rendering speed across test views ($y$-axis). Performance is measured at standard test set resolution.}
    \label{fig:fps_vs_prims}
\end{figure}

Our method, while generally faster than all baselines, also achieves an improvement in terms of quality compared to RadFoam~\cite{govindarajan2025radiant}, which is the most similar baseline to ours. While 3DGS~\cite{kerbl20233d} achieves higher quality, it suffers from popping artifacts and generally slower rendering speeds. EVER~\cite{mai2024ever} and SVRaster~\cite{sun2025sparse} have accurate volumetric rendering and high quality, but suffer from even slower rendering speeds and very limited portability, especially compared to ours. To further investigate performance, we show in Figure~\ref{fig:fps_vs_prims} that our method clearly defines the pareto front in terms of performance per primitive across a wide range of baselines.

%When compared to the broader state-of-the-art, our method does not yet surpass the highest fidelity non-mesh alternatives.
% Our representation lacks natural resilience against floaters and may benefit from additional regularizers, such as SVRaster~\cite{sun2025sparse}.
%Nevertheless, our method compensates for this marginal gap in visual quality with significantly higher framerates and greater flexibility.

%Our experiments show that in general our method achieves the highest rendering speeds among all compared methods on the Mip-NeRF 360 dataset, and achieves parity with 3DGS on the Tanks\&Temples + DeepBlending dataset\gk{parity in terms of FPS, that is a bit missleading}.
\begin{table*}[t!]
    \centering
    \small
    % \resizebox{\linewidth}{!}{
    % \large
        \begin{tabular}{l|rrr|rrr|rrr}
                                           & \multicolumn{3}{c|}{mip-NeRF 360 Outdoor~\cite{barron2022mipnerf360}}                                                           & \multicolumn{3}{c|}{mip-NeRF 360  Indoor~\cite{barron2022mipnerf360}}                                                            & \multicolumn{3}{c}{T\&T~\cite{knapitsch2017tanks}+DB~\cite{hedman2018deep}}                        \\
                                           \cline{2-10}
                                           & \multicolumn{1}{l}{PSNR $\uparrow$}     & \multicolumn{1}{l}{SSIM $\uparrow$} & \multicolumn{1}{l|}{LPIPS $\downarrow$} & \multicolumn{1}{l}{PSNR $\uparrow$}     & \multicolumn{1}{l}{SSIM $\uparrow$} & \multicolumn{1}{l|}{LPIPS $\downarrow$} & \multicolumn{1}{l}{PSNR $\uparrow$}     & \multicolumn{1}{l}{SSIM $\uparrow$} & \multicolumn{1}{l}{LPIPS $\downarrow$} \\
3DGS~\cite{kerbl20233d}                    & \cellcolor[HTML]{FFC4B3}24.63           & \cellcolor[HTML]{FFD2B3}0.729        & \cellcolor[HTML]{FFDDB3}0.271            & \cellcolor[HTML]{FFB3B3}31.05           & \cellcolor[HTML]{FFD1B3}0.924        & \cellcolor[HTML]{FFE3B3}0.238            & \cellcolor[HTML]{FFB3B3}26.65           & \cellcolor[HTML]{FFD8B3}0.876        & \cellcolor[HTML]{FFDAB3}0.263           \\
EVER~\cite{mai2024ever}                    & \cellcolor[HTML]{FFB3B3}24.73           & \cellcolor[HTML]{FFB3B3}0.744        & \cellcolor[HTML]{FFB9B3}0.239            & \cellcolor[HTML]{FFBCB3}30.98           & \cellcolor[HTML]{FFC3B3}0.926        & \cellcolor[HTML]{FFD1B3}0.227            & \cellcolor[HTML]{FFDCB4}26.09           & \cellcolor[HTML]{FFB3B3}0.889        & \cellcolor[HTML]{FFBBB3}0.234           \\
Triangle Splatting~\cite{held2025triangle} & \cellcolor[HTML]{FFEFB4}24.16           & \cellcolor[HTML]{FFDCB4}0.720        & \cellcolor[HTML]{FFC7B3}0.248            & \cellcolor[HTML]{FFD2B3}30.79           & \cellcolor[HTML]{FFDBB4}0.923        & \cellcolor[HTML]{FFB8B3}0.207            & \cellcolor[HTML]{FFC3B3}26.59           & \cellcolor[HTML]{FFDAB4}0.875        & \cellcolor[HTML]{FFC9B3}0.243           \\
SVRaster~\cite{sun2025sparse}              & \cellcolor[HTML]{FFB9B3}24.70           & \cellcolor[HTML]{FFBFB3}0.738        & \cellcolor[HTML]{FFB3B3}0.234            & \cellcolor[HTML]{FFECB4}30.67           & \cellcolor[HTML]{FFB3B3}0.927        & \cellcolor[HTML]{FFB3B3}0.203            & \cellcolor[HTML]{FFD1B3}26.52           & \cellcolor[HTML]{FFDAB4}0.874        & \cellcolor[HTML]{FFB3B3}0.229           \\
RadFoam~\cite{govindarajan2025radiant}     & \cellcolor[HTML]{FFFFB4}23.90           & \cellcolor[HTML]{FFFFB4}0.663        & \cellcolor[HTML]{FFFFB4}0.366            & \cellcolor[HTML]{FFF0B4}30.66           & \cellcolor[HTML]{FFFFB4}0.906        & \cellcolor[HTML]{FFFCB3}0.251            & \cellcolor[HTML]{FFFFB4}19.20           & \cellcolor[HTML]{FFFFB4}0.590        & \cellcolor[HTML]{FFFFB4}0.543           \\
Our Model                                  & \cellcolor[HTML]{FFE1B4}24.38           & \cellcolor[HTML]{FFDCB4}0.721        & \cellcolor[HTML]{FFE4B3}0.292            & \cellcolor[HTML]{FFFFB4}30.61           & \cellcolor[HTML]{FFE1B4}0.920        & \cellcolor[HTML]{FFFFB4}0.252            & \cellcolor[HTML]{FFDAB4}26.45           & \cellcolor[HTML]{FFCFB3}0.879        & \cellcolor[HTML]{FFDDB3}0.286           \\ \hline
                                           & \multicolumn{1}{l}{GPU-hr $\downarrow$} & \multicolumn{1}{l}{\#Prim.}    & \multicolumn{1}{l|}{FPS $\uparrow$}     & \multicolumn{1}{l}{GPU-hr $\downarrow$} & \multicolumn{1}{l}{\# Prim.}    & \multicolumn{1}{l|}{FPS $\uparrow$}     & \multicolumn{1}{l}{GPU-hr $\downarrow$} & \multicolumn{1}{l}{\# Prim.}    & \multicolumn{1}{l}{FPS $\uparrow$}     \\ \hline
3DGS~\cite{kerbl20233d}                    & \cellcolor[HTML]{FFD0B3}0.609           & \cellcolor[HTML]{FFEEB3}7.4 M       & \cellcolor[HTML]{FFE0B4}145             & \cellcolor[HTML]{FFC7B3}0.451           & \cellcolor[HTML]{FFB3B3}1.5 M       & \cellcolor[HTML]{FFC1B3}251             & \cellcolor[HTML]{FFC2B3}0.336           & \cellcolor[HTML]{FFDAB3}5.0 M       & \cellcolor[HTML]{FFB3B3}535            \\
EVER~\cite{mai2024ever}                    & \cellcolor[HTML]{FFDCB3}1.077           & \cellcolor[HTML]{FFEEB3}7.4 M       & \cellcolor[HTML]{FFFFB4}36              & \cellcolor[HTML]{FFDDB3}1.005           & \cellcolor[HTML]{FFB3B3}1.5 M       & \cellcolor[HTML]{FFFFB4}66              & \cellcolor[HTML]{FFDDB3}0.861           & \cellcolor[HTML]{FFDAB3}5.0 M       & \cellcolor[HTML]{FFFFB4}41             \\
Triangle Splatting~\cite{held2025triangle} & \cellcolor[HTML]{FFD0B3}0.612           & \cellcolor[HTML]{FFD8B3}4.7 M       & \cellcolor[HTML]{FFF1B4}87              & \cellcolor[HTML]{FFCEB3}0.558           & \cellcolor[HTML]{FFC1B3}2.3 M       & \cellcolor[HTML]{FFDFB4}137             & \cellcolor[HTML]{FFC0B3}0.313           & \cellcolor[HTML]{FFB3B3}2.2 M       & \cellcolor[HTML]{FFEBB4}200            \\
SVRaster~\cite{sun2025sparse}              & \cellcolor[HTML]{FFB5B3}0.194           & \cellcolor[HTML]{FFFFB4}9.4 M       & \cellcolor[HTML]{FFCDB3}193             & \cellcolor[HTML]{FFB8B3}0.245           & \cellcolor[HTML]{FFFFB4}8.2 M       & \cellcolor[HTML]{FFE0B4}137             & \cellcolor[HTML]{FFB3B3}0.160           & \cellcolor[HTML]{FFFFB4}7.6 M       & \cellcolor[HTML]{FFDBB4}327            \\
RadFoam~\cite{govindarajan2025radiant}     & \cellcolor[HTML]{FFE1B3}1.587           & \cellcolor[HTML]{FFC8B3}3.3 M       & \cellcolor[HTML]{FFB7B3}234             & \cellcolor[HTML]{FFE5B3}1.640           & \cellcolor[HTML]{FFCDB3}2.9 M       & \cellcolor[HTML]{FFD7B3}162             & \cellcolor[HTML]{FFFFB4}2.816           & \cellcolor[HTML]{FFB5B3}2.4 M       & \cellcolor[HTML]{FFD7B3}353            \\
Our Model                                  & \cellcolor[HTML]{FFFFB4}4.515           & \cellcolor[HTML]{FFFDB3}9.3 M       & \cellcolor[HTML]{FFB3B3}240             & \cellcolor[HTML]{FFFFB4}3.457           & \cellcolor[HTML]{FFDEB3}4.2 M       & \cellcolor[HTML]{FFB3B3}384             & \cellcolor[HTML]{FFFAB3}2.575           & \cellcolor[HTML]{FFD7B3}4.8 M       & \cellcolor[HTML]{FFBFB3}475           
\end{tabular}
    % }
    % \resizebox{\linewidth}{!}{
    % \input{tables/results}
    % }
    \vspace{-1em}
    \caption{
    Results on the 5 outdoor scenes and 4 indoor scenes from the mip-NeRF 360 dataset~\cite{barron2022mipnerf360},
    % 4 large scenes from Zip-NeRF~\cite{barron2023zipnerf} datasets,
    and 4 scenes from a combined DeepBlending~\cite{hedman2018deep} and Tanks\&Temples~\cite{knapitsch2017tanks} dataset.
    ``GPU-hr'' indicates the number of GPU hours required to train a model on an NVIDIA RTX4090. 
    FPS results are computed at the test resolution of each dataset.
    Note that our FPS results include time to display the results to the frame buffer and indicate actual performance in use, while all others capture only render time.
    }
    \label{tab:main_results}
\end{table*}

Another important metric that ensures practicality is VRAM consumption. Radiant Foam has high VRAM requirements; when running it on an RTX 4090 (24GB), we observed that it runs out of memory at $\sim$3.3 million cells. In contrast, our method supports $\sim$15 million tetrahedra on the same hardware.

Our optimization also proves to be more robust than Radiant Foam. In Tanks\&Temples+DeepBlending dataset RadFoam fails to reconstruct them in a catastrophic way, while our method achieves competitive results with all the other baselines. We hypothesize that the RadFoam tracing algorithm fails because of numerical stability issues, as a single incorrect intersection, such as on a nearly parallel face, can cause a ray to diverge.

\begin{table}[h]
\small
\begin{tabular}{l|ccc}
& \multicolumn{1}{l}{PSNR $\uparrow$} & \multicolumn{1}{l}{SSIM $\uparrow$} & \multicolumn{1}{l}{LPIPS $\downarrow$} \\ \hline
No Total Var Splitting  & \cellcolor[HTML]{FFDEB4}27.59       & \cellcolor[HTML]{FFD9B3}0.825        & \cellcolor[HTML]{FFD9B3}0.312           \\
No SSIM Splitting & \cellcolor[HTML]{FFFFB4}26.92       & \cellcolor[HTML]{FFFFB4}0.751        & \cellcolor[HTML]{FFFFB4}0.417           \\ \hline
Constant Color       & \cellcolor[HTML]{FFD9B3}27.68       & \cellcolor[HTML]{FFE0B4}0.813        & \cellcolor[HTML]{FFDDB3}0.324           \\ \hline
No Instant-NGP                 & \cellcolor[HTML]{FFEFB4}27.25       & \cellcolor[HTML]{FFEDB4}0.787        & \cellcolor[HTML]{FFEAB3}0.360           \\
No Downweighting         & \cellcolor[HTML]{FFD9B3}27.68       & \cellcolor[HTML]{FFD6B3}0.826        & \cellcolor[HTML]{FFC7B3}0.307           \\
No Centroid         & \cellcolor[HTML]{FFC4B3}27.79       & \cellcolor[HTML]{FFD6B3}0.826        & \cellcolor[HTML]{FFD1B3}0.310           \\
Our Method           & \cellcolor[HTML]{FFB3B3}27.89       & \cellcolor[HTML]{FFB3B3}0.830        & \cellcolor[HTML]{FFB3B3}0.301          
\end{tabular}
\vspace{-1em}
\caption{We evaluate the impact of different splitting criteria, the benefit of linear color ramps (versus constant color), and various methods for assigning attributes to primitives.
}
\label{tab:ablations}
\end{table}

\begin{figure}
    \centering
    \includegraphics[width=\linewidth]{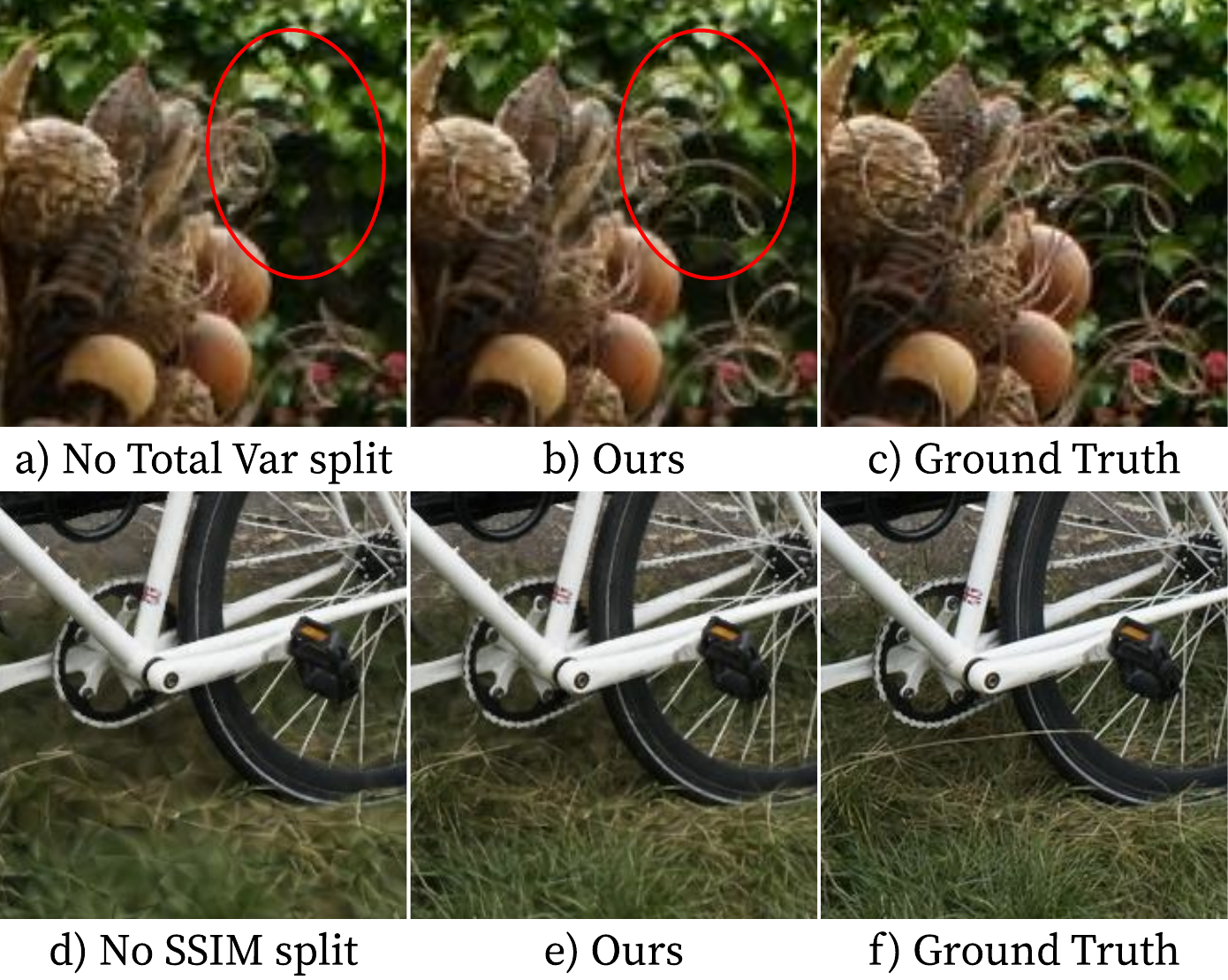}
    \vspace{-1em}
    \caption{Qualitative results of our ablation over our splitting methods. SSIM splitting handles texture, while total variance splitting handles thin structures. Red circles highlight missing detail}
    \label{fig:splitting_ablation}
\end{figure}

In Table~\ref{tab:ablations} we present our ablation study, run on the \textit{bicycle} and \textit{room} scenes, from Mip-NeRF360. We investigate different approaches for splitting and colorizing primitives, and different Instant-NGP querying approaches.
We see that omitting the SSIM based variance based densification (``SSIM Splitting") decreases performance significantly. Omitting the total dataset variance based densification (``Total Var Splitting") decreases performance marginally, and qualitatively mostly affects thin structures, shown in Fig.~\ref{fig:splitting_ablation}.
Assigning each tetrahedron a single color (``Constant Color'') instead of using a linear color results in a marginal decrease in quality.
Ablating the Instant-NGP (``No Instant-NGP'') and directly assigning properties to the vertices works poorly due to topology flips, as expected.
Ablating the downweighting used after in interpolation (``No Downweighting") hurts slightly.
(``No Centroid") is a test where we used the circumcenter instead of the centroid when querying the iNGP to get the value of each tet. As described in Sec.~\ref{sec:model}, the circumcenter is mathematically smooth but has worse numerical conditioning, a tradeoff that this experiment shows to be disadvantageous.

\begin{figure}[h]
    \centering
    \includegraphics[width=\linewidth]{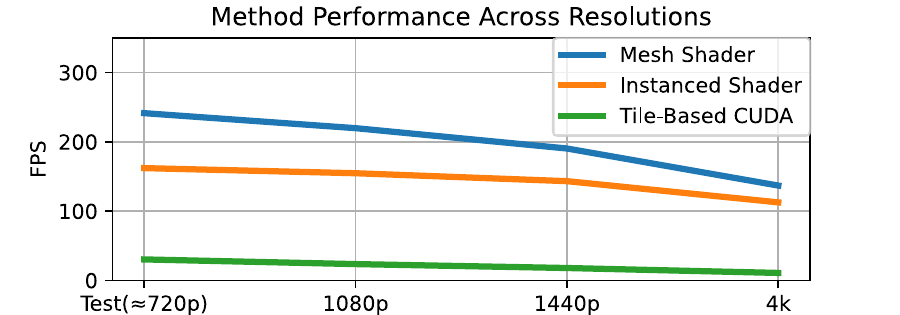}
    \caption{We plot the frames per second across a wide range of resolutions that are common in modern hardware. We show how rendering tetrahedra with a tile-based approach similar to 3DGS is significantly slower than hardware-based rasterizers. We also show how  mesh shaders improve performance compared to instance-based rendering by reducing the number of duplicate memory loads.
    }
    \label{fig:speed_ablation}
\end{figure}

In Figure~\ref{fig:speed_ablation} we analyze different shader implementations to measure the impact of each choice. 
``Tile-based'' is a naive tile-based renderer that we adapted from the 3DGS SlangTorch codebase, and ``Instanced'' is a hardware triangle rasterizer with duplicated loads.
We see that switching from the tile-based renderer to the instanced hardware triangle rasterizer that utilizes vertex and fragment shaders yields a significant performance boost.
As detailed in Sec.~\ref{sec:model}, the standard rasterization pipeline incurs redundant primitive loads in the vertex shader.
We see that using our mesh shader pipeline eliminates this redundancy and achieves another significant performance gain.

\subsection{Capabilities}

\begin{figure}
    \centering
    \includegraphics[width=\linewidth]{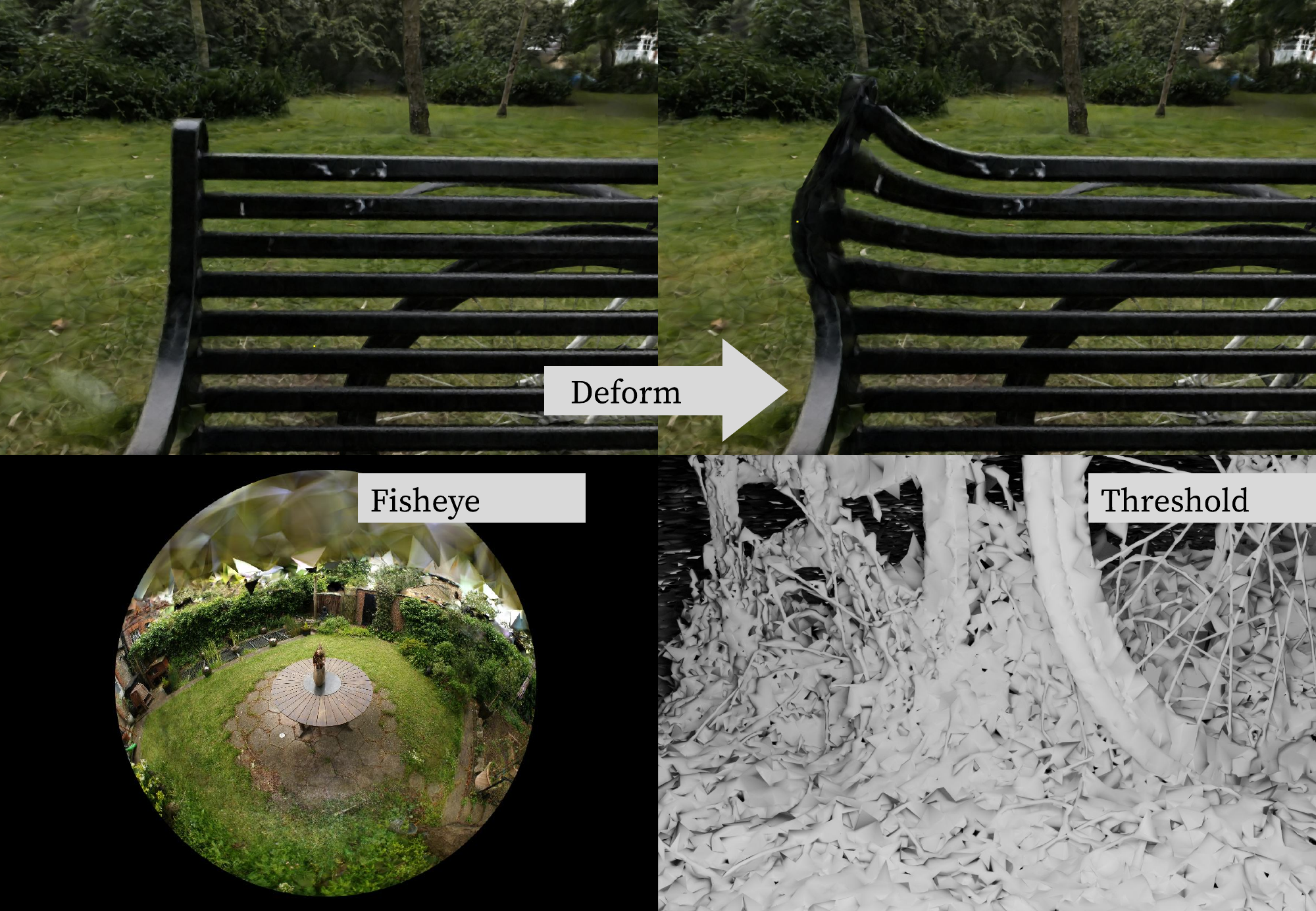}
    \caption{Radiance meshes enable a variety of useful capabilities. 
    Bottom left: our splatting method supports fisheye during training because power sorting supports any camera where rays share a common origin.
    Bottom right: Blender rendering of a surface mesh of the \scenename{bicycle} scene obtained by thresholding the scene.
    Top right: To simulate physics in realtime, we apply xPBD~\cite{macklin2016xpbd} to the radiance mesh vertices.}
    \label{fig:capabilities}
\end{figure}

Our model enables a variety of qualitative capabilities, all of which are shown in Figure~\ref{fig:capabilities}.

\myparagraph{Lens Models \& Ray Tracing}
Because our representation is highly general, we were able to straightforwardly implement a ray-tracer based on OptiX, by leveraging the transparency method from 3DGRT~\cite{3dgrt2024}. Crucially, our tetrahedral mesh allows us to utilize built-in hardware triangle intersection. We achieve 84 FPS on the MipNeRF 360 outdoor scenes and 190 FPS on the indoor scenes, at test resolution (roughly 720p on outdoor scenes, less than 1080p on indoor scenes). At comparable primitive counts, we achieve a 17\% speed increase relative to Radiant Foam's ray tracer. This enables us to train on arbitrary lenses.

\myparagraph{Surface mesh extraction}
Our volumetric representation can be converted into a surface mesh using a straightforward procedure: We measure the peak color contribution of each tetrahedron across all pixels and all training views. We use this quantity to then threshold the radiance mesh by removing the tetrahedrons with a smaller peak color contribution than $0.1$. We then identify the connected components of all tetrahedra and construct a surface mesh for each one.

\myparagraph{Deformation and Simulation}
Finite element analysis and various simulation techniques like Position Based Dynamics (PBD)~\cite{muller2007position, macklin2016xpbd} are proven to be very useful and practical in interactive graphics applications. We demonstrate in Fig.~\ref{fig:capabilities} that a radiance mesh can directly use PBD and xPDB techniques to create interactive physical simulators, we also provide recorded interactive sessions in our supplemental video. This is a direct consequence of the nature of radiance meshes, as we are using the primal form of the Voronoi diagram --- a simple Tetrahedral mesh. Previous work~\cite{govindarajan2025radiant} that use the dual form of a Voronoi diagram only store, optimize, and render the generators of each cell. Animating these generator sites would cause non-trivial topological changes to the cell shapes. 

%However, the representation used by Radiant Foam~\cite{govindarajan2025radiant} is in its dual form, which lacks explicit vertices \jb{this reads like a complete non-sequitor, why are you bringing up Radiant Foam?}; it only stores, optimizes, and renders the generators of each cell. Animating these generator sites causes non-trivial topological changes to the cell shapes. Furthermore, switching from the dual to the primal form for simulation would break the Radiant Foam rendering method and introduce tens of millions of polyhedral faces, which are expensive to render as they require triangulation.

%In contrast, our method optimizes and renders the primal form directly. This allows us to apply methods like PBD and xPBD natively to our \longnamesingular{} and view the results using our standard rendering techniques, as shown on the right of Fig.~\ref{fig:capabilities}.

%\jb{You should rewrite this whole section so that you first spend many words explaining how your model works well for this, and then you should briefly explain why prior work doesn't do this, instead of the opposite.}

% \begin{figure*}
%     \centering
%     \includegraphics[width=\linewidth]{figures/FPS.pdf}
%     \caption{Mesh rendering is not available on the NVIDIA A5000. As a result, we fallback to instanced rendering. A5000 results are not up to date yet.}
%     \label{fig:fps}
% \end{figure*}

\section{Conclusion}

We have presented radiance meshes, a novel hybrid radiance field representation that leverages the Delaunay triangulation to surpass the speed and flexibility of fast particle-based approaches, while still providing the physical consistency, and the lack of temporal artifacts of much slower neural field approaches. To accomplish this, we have presented a novel way to parameterize the attributes of the tetrahedra yielded by a Delaunay triangulation using an Instant-NGP. We additionally presented techniques for tetrahedral densification and hardware accelerated tetrahedral rendering.
We have shown that this flexible backwards-compatible representation enables a variety of standard editing applications and physics-based simulation approaches, in addition to supporting arbitrary lens distortions.
We believe that our approach represents a compelling combination of highly desirable properties for radiance field models, and we hope that this work will bring the community closer to widespread adoption of radiance field techniques in user-facing computer graphics applications.

\iftoggle{iccvfinal}{
\myparagraph{Acknowledgements}
We would like to thank the GraphDeco team at Inria, for providing a stimulating research environment during the final three months of this work.
AM is funded in part by the USACE Engineer Research and Development Center Cooperative Agreement W9132T-22-2-0014. AM conducted the final steps of research at GraphDeco@Inria.
}{}

{
    \small
    \bibliographystyle{ieeenat_fullname}
    \bibliography{main}

\begin{thebibliography}{68}
\providecommand{\natexlab}[1]{#1}
\providecommand{\url}[1]{\texttt{#1}}
\expandafter\ifx\csname urlstyle\endcsname\relax
  \providecommand{\doi}[1]{doi: #1}\else
  \providecommand{\doi}{doi: \begingroup \urlstyle{rm}\Url}\fi

\bibitem[Aliev et~al.(2020)Aliev, Sevastopolsky, Kolos, Ulyanov, and Lempitsky]{aliev2020neural}
Kara-Ali Aliev, Artem Sevastopolsky, Maria Kolos, Dmitry Ulyanov, and Victor Lempitsky.
\newblock {Neural Point-Based Graphics}.
\newblock \emph{ECCV}, 2020.

\bibitem[Bangaru et~al.(2023)Bangaru, Wu, Li, Munkberg, Bernstein, Ragan-Kelley, Durand, Lefohn, and He]{bangaru2023slangd}
Sai Bangaru, Lifan Wu, Tzu-Mao Li, Jacob Munkberg, Gilbert Bernstein, Jonathan Ragan-Kelley, Fredo Durand, Aaron Lefohn, and Yong He.
\newblock {SLANG.D: Fast, Modular and Differentiable Shader Programming}.
\newblock \emph{SIGGRAPH Asia}, 2023.

\bibitem[Barron et~al.(2021)Barron, Mildenhall, Tancik, Hedman, Martin-Brualla, and Srinivasan]{barron2021mip}
Jonathan~T. Barron, Ben Mildenhall, Matthew Tancik, Peter Hedman, Ricardo Martin-Brualla, and Pratul~P. Srinivasan.
\newblock {Mip-NeRF: A Multiscale Representation for Anti-Aliasing Neural Radiance Fields}.
\newblock \emph{ICCV}, 2021.

\bibitem[Barron et~al.(2022)Barron, Mildenhall, Verbin, Srinivasan, and Hedman]{barron2022mipnerf360}
Jonathan~T. Barron, Ben Mildenhall, Dor Verbin, Pratul~P. Srinivasan, and Peter Hedman.
\newblock {Mip-NeRF 360: Unbounded Anti-Aliased Neural Radiance Fields}.
\newblock \emph{CVPR}, 2022.

\bibitem[Barron et~al.(2023)Barron, Mildenhall, Verbin, Srinivasan, and Hedman]{barron2023zipnerf}
Jonathan~T. Barron, Ben Mildenhall, Dor Verbin, Pratul~P. Srinivasan, and Peter Hedman.
\newblock {Zip-NeRF: Anti-Aliased Grid-Based Neural Radiance Fields}.
\newblock \emph{ICCV}, 2023.

\bibitem[Chan et~al.(2022)Chan, Lin, Chan, Nagano, Pan, De~Mello, Gallo, Guibas, Tremblay, Khamis, et~al.]{chan2022efficient}
Eric~R Chan, Connor~Z Lin, Matthew~A Chan, Koki Nagano, Boxiao Pan, Shalini De~Mello, Orazio Gallo, Leonidas~J Guibas, Jonathan Tremblay, Sameh Khamis, et~al.
\newblock {Efficient Geometry-Aware 3D Generative Adversarial Networks}.
\newblock \emph{CVPR}, 2022.

\bibitem[Chandrasekhar(1960)]{chandrasekhar1960radiative}
Subrahmanyan Chandrasekhar.
\newblock \emph{{Radiative Transfer}}.
\newblock Courier Corporation, 1960.

\bibitem[Chen et~al.(2024)Chen, Li, Ye, Wang, Xie, Zhai, Wang, Liu, Bao, and Zhang]{Chen_2024}
Danpeng Chen, Hai Li, Weicai Ye, Yifan Wang, Weijian Xie, Shangjin Zhai, Nan Wang, Haomin Liu, Hujun Bao, and Guofeng Zhang.
\newblock {PGSR: Planar-based Gaussian Splatting for Efficient and High-Fidelity Surface Reconstruction}.
\newblock \emph{Visualization and Computer Graphics}, 2024.

\bibitem[Chen et~al.(2023)Chen, Funkhouser, Hedman, and Tagliasacchi]{chen2023mobilenerf}
Zhiqin Chen, Thomas Funkhouser, Peter Hedman, and Andrea Tagliasacchi.
\newblock {MobileNeRF: Exploiting the Polygon Rasterization Pipeline for Efficient Neural Field Rendering on Mobile Architectures}.
\newblock \emph{CVPR}, 2023.

\bibitem[Choi et~al.(2024)Choi, Lee, Lee, Kwon, and Manocha]{choi2024meshgs}
Jaehoon Choi, Yonghan Lee, Hyungtae Lee, Heesung Kwon, and Dinesh Manocha.
\newblock {MeshGS: Adaptive Mesh-Aligned Gaussian Splatting for High-Quality Rendering}.
\newblock \emph{ACCV}, 2024.

\bibitem[Cyrus and Beck(1978)]{cyrus1978}
Mike Cyrus and Jay Beck.
\newblock {Generalized Two- and Three-Dimensional Clipping}.
\newblock \emph{Computers \& Graphics}, 1978.

\bibitem[Delaunay(1934)]{delaunay1934sphere}
B Delaunay.
\newblock {Sur la Sph{\`e}re Vide}.
\newblock \emph{Bulletin de l’Acad{\'e}mie des Sciences de l’URSS, Classe des sciences math{\'e}matiques et naturelles}, 1934.

\bibitem[Edelsbrunner(1989)]{edelsbrunner1989acyclicity}
Herbert Edelsbrunner.
\newblock An acyclicity theorem for cell complexes in d dimensions.
\newblock \emph{Symposium on Computational geometry}, 1989.

\bibitem[Elsner et~al.(2023)Elsner, Czech, Berger, Selman, Lim, and Kobbelt]{elsner2023adaptive}
Tim Elsner, Victor Czech, Julia Berger, Zain Selman, Isaak Lim, and Leif Kobbelt.
\newblock {Adaptive Voronoi NeRFs}.
\newblock \emph{arXiv:2303.16001}, 2023.

\bibitem[Fridovich-Keil et~al.(2022)Fridovich-Keil, Yu, Tancik, Chen, Recht, and Kanazawa]{fridovich2022plenoxels}
Sara Fridovich-Keil, Alex Yu, Matthew Tancik, Qinhong Chen, Benjamin Recht, and Angjoo Kanazawa.
\newblock {Plenoxels: Radiance fields without neural networks}.
\newblock \emph{CVPR}, 2022.

\bibitem[Gao et~al.(2024)Gao, Yang, Zhang, Sun, Yuan, Fu, and Lai]{gao2024real}
Lin Gao, Jie Yang, Bo-tao Zhang, Jia-mu Sun, Yu-jie Yuan, Hongbo Fu, and Yu-kun Lai.
\newblock {Real-time Large-scale Deformation of Gaussian Splatting}.
\newblock \emph{ACM Transactions on Graphics (TOG)}, 2024.

\bibitem[Gao et~al.(2025)Gao, Li, Zhuang, Zhang, Hu, Zhang, Yao, Shan, and Quan]{gao2025mani}
Xiangjun Gao, Xiaoyu Li, Yiyu Zhuang, Qi Zhang, Wenbo Hu, Chaopeng Zhang, Yao Yao, Ying Shan, and Long Quan.
\newblock {Mani-GS: Gaussian Splatting Manipulation with Triangular Mesh}.
\newblock \emph{CVPR}, 2025.

\bibitem[Georgii and Westermann(2006)]{georgii2006generic}
Joachim Georgii and Rudiger Westermann.
\newblock {A Generic and Scalable Pipeline for GPU Tetrahedral Grid Rendering}.
\newblock \emph{IEEE Transactions on Visualization and Computer Graphics}, 2006.

\bibitem[Govindarajan et~al.(2025)Govindarajan, Rebain, Yi, and Tagliasacchi]{govindarajan2025radiant}
Shrisudhan Govindarajan, Daniel Rebain, Kwang~Moo Yi, and Andrea Tagliasacchi.
\newblock {Radiant Foam: Real-Time Differentiable Ray Tracing}.
\newblock \emph{ICCV}, 2025.

\bibitem[Gu{\'e}don et~al.(2025)Gu{\'e}don, Gomez, Maruani, Gong, Drettakis, and Ovsjanikov]{guedon2025milo}
Antoine Gu{\'e}don, Diego Gomez, Nissim Maruani, Bingchen Gong, George Drettakis, and Maks Ovsjanikov.
\newblock {MILo: Mesh-In-the-Loop Gaussian Splatting for Detailed and Efficient Surface Reconstruction}.
\newblock \emph{SIGGRAPH Asia}, 2025.

\bibitem[Guo et~al.(2024)Guo, Wang, He, and Matusik]{guo2024tetsphere}
Minghao Guo, Bohan Wang, Kaiming He, and Wojciech Matusik.
\newblock {TetSphere Splatting: Representing High-Quality Geometry with Lagrangian Volumetric Meshes}.
\newblock \emph{arXiv:2405.20283}, 2024.

\bibitem[Hamdi et~al.(2024)Hamdi, Melas-Kyriazi, Mai, Qian, Liu, Vondrick, Ghanem, and Vedaldi]{hamdi2024ges}
Abdullah Hamdi, Luke Melas-Kyriazi, Jinjie Mai, Guocheng Qian, Ruoshi Liu, Carl Vondrick, Bernard Ghanem, and Andrea Vedaldi.
\newblock {GES: Generalized Exponential Splatting for Efficient Radiance Field Rendering}.
\newblock \emph{CVPR}, 2024.

\bibitem[Hedman et~al.(2018)Hedman, Philip, Price, Frahm, Drettakis, and Brostow]{hedman2018deep}
Peter Hedman, Julien Philip, True Price, Jan-Michael Frahm, George Drettakis, and Gabriel Brostow.
\newblock Deep blending for free-viewpoint image-based rendering.
\newblock \emph{ACM Transactions on Graphics (TOG)}, 2018.

\bibitem[Held et~al.(2025{\natexlab{a}})Held, Vandeghen, Deliege, Hamdi, Giancola, Cioppa, Vedaldi, Ghanem, Tagliasacchi, and Van~Droogenbroeck]{held2025triangle}
Jan Held, Renaud Vandeghen, Adrien Deliege, Abdullah Hamdi, Silvio Giancola, Anthony Cioppa, Andrea Vedaldi, Bernard Ghanem, Andrea Tagliasacchi, and Marc Van~Droogenbroeck.
\newblock {Triangle Splatting for Real-Time Radiance Field Rendering}.
\newblock \emph{arXiv:2505.19175}, 2025{\natexlab{a}}.

\bibitem[Held et~al.(2025{\natexlab{b}})Held, Vandeghen, Hamdi, Deliege, Cioppa, Giancola, Vedaldi, Ghanem, and Van~Droogenbroeck]{held20253d}
Jan Held, Renaud Vandeghen, Abdullah Hamdi, Adrien Deliege, Anthony Cioppa, Silvio Giancola, Andrea Vedaldi, Bernard Ghanem, and Marc Van~Droogenbroeck.
\newblock {3D Convex Splatting: Radiance Field Rendering with 3D Smooth Convexes}.
\newblock \emph{CVPR}, 2025{\natexlab{b}}.

\bibitem[Huang et~al.(2024)Huang, Yu, Chen, Geiger, and Gao]{huang20242d}
Binbin Huang, Zehao Yu, Anpei Chen, Andreas Geiger, and Shenghua Gao.
\newblock {2D Gaussian Splatting for Geometrically Accurate Radiance Fields}.
\newblock \emph{SIGGRAPH 2024}, 2024.

\bibitem[Huang et~al.(2025)Huang, Lin, Sun, Yang, Lyu, Cao, and Qi]{huang2025deformable}
Yi-Hua Huang, Ming-Xian Lin, Yang-Tian Sun, Ziyi Yang, Xiaoyang Lyu, Yan-Pei Cao, and Xiaojuan Qi.
\newblock {Deformable Radial Kernel Splatting}.
\newblock \emph{CVPR}, 2025.

\bibitem[Hurtado et~al.(1996)Hurtado, Noy, and Urrutia]{hurtado1996flipping}
Ferran Hurtado, Marc Noy, and Jorge Urrutia.
\newblock {Flipping Edges in Triangulations}.
\newblock \emph{Symposium on Computational Geometry}, 1996.

\bibitem[Karasick et~al.(1997)Karasick, Lieber, Nackman, and Rajan]{karasick1997visualization}
Michael~S Karasick, Derek Lieber, Lee~R. Nackman, and VT Rajan.
\newblock {Visualization of Three-Dimensional Delaunay Meshes}.
\newblock \emph{Algorithmica}, 1997.

\bibitem[Kato et~al.(2018)Kato, Ushiku, and Harada]{kato2018neural}
Hiroharu Kato, Yoshitaka Ushiku, and Tatsuya Harada.
\newblock {Neural 3D Mesh Renderer}.
\newblock \emph{CVPR}, 2018.

\bibitem[Kerbl et~al.(2023)Kerbl, Kopanas, Leimk{\"u}hler, and Drettakis]{kerbl20233d}
Bernhard Kerbl, Georgios Kopanas, Thomas Leimk{\"u}hler, and George Drettakis.
\newblock {3D Gaussian Splatting for Real-Time Radiance Field Rendering}.
\newblock \emph{ACM Transactions on Graphics (TOG)}, 2023.

\bibitem[Knapitsch et~al.(2017)Knapitsch, Park, Zhou, and Koltun]{knapitsch2017tanks}
Arno Knapitsch, Jaesik Park, Qian-Yi Zhou, and Vladlen Koltun.
\newblock {Tanks and Temples: Benchmarking Large-Scale Scene Reconstruction}.
\newblock \emph{ACM Transactions on Graphics (TOG)}, 2017.

\bibitem[Lassner and Zollhofer(2021)]{lassner2021pulsar}
Christoph Lassner and Michael Zollhofer.
\newblock {Pulsar: Efficient Sphere-Based Neural Rendering}.
\newblock \emph{CVPR}, 2021.

\bibitem[Li et~al.(2024)Li, Liu, Sznaier, and Camps]{li20243d}
Haolin Li, Jinyang Liu, Mario Sznaier, and Octavia Camps.
\newblock {3D-HGS: 3D Half-Gaussian Splatting}.
\newblock \emph{arXiv:2406.02720}, 2024.

\bibitem[Li et~al.(2018)Li, Aittala, Durand, and Lehtinen]{li2018differentiable}
Tzu-Mao Li, Miika Aittala, Fr{\'e}do Durand, and Jaakko Lehtinen.
\newblock Differentiable monte carlo ray tracing through edge sampling.
\newblock \emph{ACM Transactions on Graphics (TOG)}, 2018.

\bibitem[Lin et~al.(2024)Lin, Xiang, Kennedy, and Li]{lin2024direct}
Ancheng Lin, Yusheng Xiang, Paul Kennedy, and Jun Li.
\newblock {Direct Learning of Mesh and Appearance via 3D Gaussian Splatting}.
\newblock \emph{arXiv:2405.06945}, 2024.

\bibitem[Liu et~al.(2025)Liu, Sun, Chen, Wang, and Feng]{liu2025deformable}
Rong Liu, Dylan Sun, Meida Chen, Yue Wang, and Andrew Feng.
\newblock {Deformable Beta Splatting}.
\newblock \emph{SIGGRAPH}, 2025.

\bibitem[Lombardi et~al.(2021)Lombardi, Simon, Schwartz, Zollhoefer, Sheikh, and Saragih]{lombardi2021mixture}
Stephen Lombardi, Tomas Simon, Gabriel Schwartz, Michael Zollhoefer, Yaser Sheikh, and Jason Saragih.
\newblock Mixture of volumetric primitives for efficient neural rendering.
\newblock \emph{ACM TOG}, 2021.

\bibitem[Macklin et~al.(2016)Macklin, M{\"u}ller, and Chentanez]{macklin2016xpbd}
Miles Macklin, Matthias M{\"u}ller, and Nuttapong Chentanez.
\newblock Xpbd: position-based simulation of compliant constrained dynamics.
\newblock \emph{International Conference on Motion in Games}, 2016.

\bibitem[Mai et~al.(2025)Mai, Hedman, Kopanas, Verbin, Futschik, Xu, Kuester, Barron, and Zhang]{mai2024ever}
Alexander Mai, Peter Hedman, George Kopanas, Dor Verbin, David Futschik, Qiangeng Xu, Falko Kuester, Jonathan~T Barron, and Yinda Zhang.
\newblock {EVER: Exact Volumetric Ellipsoid Rendering for Real-time View Synthesis}.
\newblock \emph{ICCV}, 2025.

\bibitem[Max et~al.(1990)Max, Hanrahan, and Crawfis]{max1990area}
Nelson Max, Pat Hanrahan, and Roger Crawfis.
\newblock {Area and Volume Coherence for Efficient Visualization of 3D Scalar Functions}.
\newblock \emph{Workshop on Volume Visualization}, 1990.

\bibitem[Meijering(1953)]{meijering1953interface}
J.~L. Meijering.
\newblock Interface area, edge length, and number of vertices in crystal aggregates with random nucleation.
\newblock \emph{Philips Research Reports}, 1953.

\bibitem[Mildenhall et~al.(2020)Mildenhall, Srinivasan, Tancik, Barron, Ramamoorthi, and Ng]{mildenhall2020nerf}
Ben Mildenhall, Pratul~P Srinivasan, Matthew Tancik, Jonathan~T Barron, Ravi Ramamoorthi, and Ren Ng.
\newblock {NeRF: Representing Scenes as Neural Radiance Fields for View Synthesis}.
\newblock \emph{ECCV}, 2020.

\bibitem[Moenne-Loccoz et~al.(2024)Moenne-Loccoz, Mirzaei, Perel, de~Lutio, Esturo, State, Fidler, Sharp, and Gojcic]{3dgrt2024}
Nicolas Moenne-Loccoz, Ashkan Mirzaei, Or Perel, Riccardo de Lutio, Janick~Martinez Esturo, Gavriel State, Sanja Fidler, Nicholas Sharp, and Zan Gojcic.
\newblock 3d gaussian ray tracing: Fast tracing of particle scenes.
\newblock \emph{SIGGRAPH Asia}, 2024.

\bibitem[M{\"u}ller et~al.(2007)M{\"u}ller, Heidelberger, Hennix, and Ratcliff]{muller2007position}
Matthias M{\"u}ller, Bruno Heidelberger, Marcus Hennix, and John Ratcliff.
\newblock Position based dynamics.
\newblock \emph{Journal of Visual Communication and Image Representation}, 2007.

\bibitem[M\"uller et~al.(2022)M\"uller, Evans, Schied, and Keller]{mueller2022instant}
Thomas M\"uller, Alex Evans, Christoph Schied, and Alexander Keller.
\newblock {Instant Neural Graphics Primitives with a Multiresolution Hash Encoding}.
\newblock \emph{ACM Transactions on Graphics (TOG)}, 2022.

\bibitem[Museth and Lombeyda(2004)]{museth2004tetsplat}
Ken Museth and Santiago Lombeyda.
\newblock {TetSplat Real-Time Rendering and Volume Clipping of Large Unstructured Tetrahedral Meshes}.
\newblock \emph{IEEE Visualization}, 2004.

\bibitem[Pachner(1991)]{pachner1991pl}
Udo Pachner.
\newblock Pl homeomorphic manifolds are equivalent by elementary shellings.
\newblock \emph{European Journal of Combinatorics}, 1991.

\bibitem[Qian et~al.(2024)Qian, Kirschstein, Schoneveld, Davoli, Giebenhain, and Nie{\ss}ner]{qian2024gaussianavatars}
Shenhan Qian, Tobias Kirschstein, Liam Schoneveld, Davide Davoli, Simon Giebenhain, and Matthias Nie{\ss}ner.
\newblock {GaussianAvatars: Photorealistic Head Avatars with Rigged 3D Gaussians}.
\newblock \emph{CVPR}, 2024.

\bibitem[Qu et~al.(2024)Qu, Li, Rahmani, Cai, and Liu]{qu2024disc}
Haoxuan Qu, Zhuoling Li, Hossein Rahmani, Yujun Cai, and Jun Liu.
\newblock {DisC-GS: Discontinuity-aware Gaussian Splatting}.
\newblock \emph{NeurIPS}, 37, 2024.

\bibitem[Radl et~al.(2024)Radl, Steiner, Parger, Weinrauch, Kerbl, and Steinberger]{radl2024stopthepop}
Lukas Radl, Michael Steiner, Mathias Parger, Alexander Weinrauch, Bernhard Kerbl, and Markus Steinberger.
\newblock {StopThePop: Sorted Gaussian Splatting for View-Consistent Real-time Rendering}, 2024.

\bibitem[Reiser et~al.(2024)Reiser, Garbin, Srinivasan, Verbin, Szeliski, Mildenhall, Barron, Hedman, and Geiger]{reiser2024binary}
Christian Reiser, Stephan Garbin, Pratul Srinivasan, Dor Verbin, Richard Szeliski, Ben Mildenhall, Jonathan Barron, Peter Hedman, and Andreas Geiger.
\newblock Binary opacity grids: Capturing fine geometric detail for mesh-based view synthesis.
\newblock \emph{ACM Transactions on Graphics (TOG)}, 2024.

\bibitem[Rota~Bul{\`o} et~al.(2024)Rota~Bul{\`o}, Porzi, and Kontschieder]{rota2024revising}
Samuel Rota~Bul{\`o}, Lorenzo Porzi, and Peter Kontschieder.
\newblock {Revising Densification in Gaussian Splatting}.
\newblock \emph{ECCV}, 2024.

\bibitem[Sch\"{o}nberger and Frahm(2016)]{schoenberger2016sfm}
Johannes~Lutz Sch\"{o}nberger and Jan-Michael Frahm.
\newblock Structure-from-motion revisited.
\newblock \emph{CVPR}, 2016.

\bibitem[Shao et~al.(2024)Shao, Wang, Li, Wang, Lin, Zhang, Fan, and Wang]{shao2024splattingavatar}
Zhijing Shao, Zhaolong Wang, Zhuang Li, Duotun Wang, Xiangru Lin, Yu Zhang, Mingming Fan, and Zeyu Wang.
\newblock {SplattingAvatar: Realistic Real-Time Human Avatars with Mesh-Embedded Gaussian Splatting}.
\newblock \emph{CVPR}, 2024.

\bibitem[Sun et~al.(2025)Sun, Choe, Loop, Ma, and Wang]{sun2025sparse}
Cheng Sun, Jaesung Choe, Charles Loop, Wei-Chiu Ma, and Yu-Chiang~Frank Wang.
\newblock {Sparse Voxels Rasterization: Real-time High-fidelity Radiance Field Rendering}.
\newblock \emph{CVPR}, 2025.

\bibitem[Tricard(2024)]{tricard2024interval}
Thibault Tricard.
\newblock Interval shading: using mesh shaders to generate shading intervals for volume rendering.
\newblock \emph{Computer Graphics and Interactive Techniques}, 2024.

\bibitem[von L{\"u}tzow and Nie{\ss}ner(2025)]{von2025linprim}
Nicolas von L{\"u}tzow and Matthias Nie{\ss}ner.
\newblock {LinPrim: Linear Primitives for Differentiable Volumetric Rendering}.
\newblock \emph{arXiv:2501.16312}, 2025.

\bibitem[Waczy{\'n}ska et~al.(2024)Waczy{\'n}ska, Borycki, Tadeja, Tabor, and Spurek]{waczynska2024games}
Joanna Waczy{\'n}ska, Piotr Borycki, S{\l}awomir Tadeja, Jacek Tabor, and Przemys{\l}aw Spurek.
\newblock {GaMeS: Mesh-Based Adapting and Modification of Gaussian Splatting}.
\newblock \emph{arXiv:2402.01459}, 2024.

\bibitem[Wang et~al.(2024)Wang, Liu, Wang, Lin, Hou, Li, Komura, and Wang]{wang2024gaussurf}
Jiepeng Wang, Yuan Liu, Peng Wang, Cheng Lin, Junhui Hou, Xin Li, Taku Komura, and Wenping Wang.
\newblock {GausSurf: Geometry-Guided 3D Gaussian Splatting for Surface Reconstruction}.
\newblock \emph{arXiv:2411.19454}, 2024.

\bibitem[Wang et~al.(2021)Wang, Liu, Liu, Theobalt, Komura, and Wang]{wang2021neus}
P Wang, L Liu, Y Liu, C Theobalt, T Komura, and WP Wang.
\newblock {NeuS: Learning Neural Implicit Surfaces by Volume Rendering for Multi-view Reconstruction}.
\newblock \emph{Neurips}, 2021.

\bibitem[Wiles et~al.(2020)Wiles, Gkioxari, Szeliski, and Johnson]{wiles2020synsin}
Olivia Wiles, Georgia Gkioxari, Richard Szeliski, and Justin Johnson.
\newblock {SynSin: End-to-end View Synthesis from a Single Image}.
\newblock \emph{CVPR}, 2020.

\bibitem[Xu et~al.(2022)Xu, Xu, Philip, Bi, Shu, Sunkavalli, and Neumann]{xu2022point}
Qiangeng Xu, Zexiang Xu, Julien Philip, Sai Bi, Zhixin Shu, Kalyan Sunkavalli, and Ulrich Neumann.
\newblock {Point-NeRF: Point-based Neural Radiance Fields}.
\newblock \emph{CVPR}, 2022.

\bibitem[Yariv et~al.(2021)Yariv, Gu, Kasten, and Lipman]{yariv2021volume}
Lior Yariv, Jiatao Gu, Yoni Kasten, and Yaron Lipman.
\newblock Volume rendering of neural implicit surfaces.
\newblock \emph{NeurIPS}, 2021.

\bibitem[Yariv et~al.(2023)Yariv, Hedman, Reiser, Verbin, Srinivasan, Szeliski, Barron, and Mildenhall]{yariv2023bakedsdf}
Lior Yariv, Peter Hedman, Christian Reiser, Dor Verbin, Pratul~P. Srinivasan, Richard Szeliski, Jonathan~T. Barron, and Ben Mildenhall.
\newblock {BakedSDF: Meshing Neural SDFs for Real-Time View Synthesis}.
\newblock \emph{SIGGRAPH}, 2023.

\bibitem[Yu et~al.(2024)Yu, Sattler, and Geiger]{yu2024gaussian}
Zehao Yu, Torsten Sattler, and Andreas Geiger.
\newblock {Gaussian Opacity Fields: Efficient Adaptive Surface Reconstruction in Unbounded Scenes}.
\newblock \emph{ACM Transactions on Graphics (ToG)}, 2024.

\bibitem[Zhang et~al.(2024)Zhang, Huang, Jiang, Zhou, Xiang, and Shen]{zhang2024quadratic}
Ziyu Zhang, Binbin Huang, Hanqing Jiang, Liyang Zhou, Xiaojun Xiang, and Shunhan Shen.
\newblock {Quadratic Gaussian Splatting for Efficient and Detailed Surface Reconstruction}.
\newblock \emph{arXiv:2411.16392}, 2024.

\bibitem[Zheng et~al.(2025)Zheng, Xue, Zarate, and Song]{zheng2025gaustar}
Chengwei Zheng, Lixin Xue, Juan Zarate, and Jie Song.
\newblock {GauSTAR: Gaussian Surface Tracking and Reconstruction}.
\newblock \emph{CVPR}, 2025.

\end{thebibliography}
}

% WARNING: do not forget to delete the supplementary pages from your submission 
\clearpage
\setcounter{page}{1}
\maketitlesupplementary

\section{Smoothness Proof Sketch}
A Delaunay triangulation can be formed by taking a triangulation and then flipping the edges of triangles that are not Delaunay (e.g.\ the edges that have a circumsphere containing a vertex)~\cite{hurtado1996flipping}. In 3D, these edge-flips are called Pachner moves~\cite{pachner1991pl}.
As discussed in Sec.~\ref{sec:query},
% during a Delaunay triangulation arbitrary edge flips may occur
during optimization we make take a valid Delaunay triangulation and then move the vertices by some small amount $\epsilon$
such that a vertex may enter the circumsphere of a neighboring tetrahedron, thereby triggering a topological flip.
% by some $\epsilon$, a flip may take place when a circumsphere from the previous triangulation now contains a vertex.
This implies that the vertex was already nearby the circumsphere boundary prior to the move. 
% For this to have happened, the offending vertex must have been close to the offending circumsphere. 
% If we form a triangle using the offending vertex and any of the vertices from the offending circumsphere, then it must form a circumsphere similar to the offending circumsphere. Therefore, the flip must have been caused by points from a similar circumsphere.
Consequently, these flips must involve sets of points that are nearly cospherical, as the geometric configurations before and after the flip share an almost identical circumsphere.

\section{Sorting Proof Sketch}
\begin{figure}[h]
    \centering
    \includegraphics[width=\linewidth]{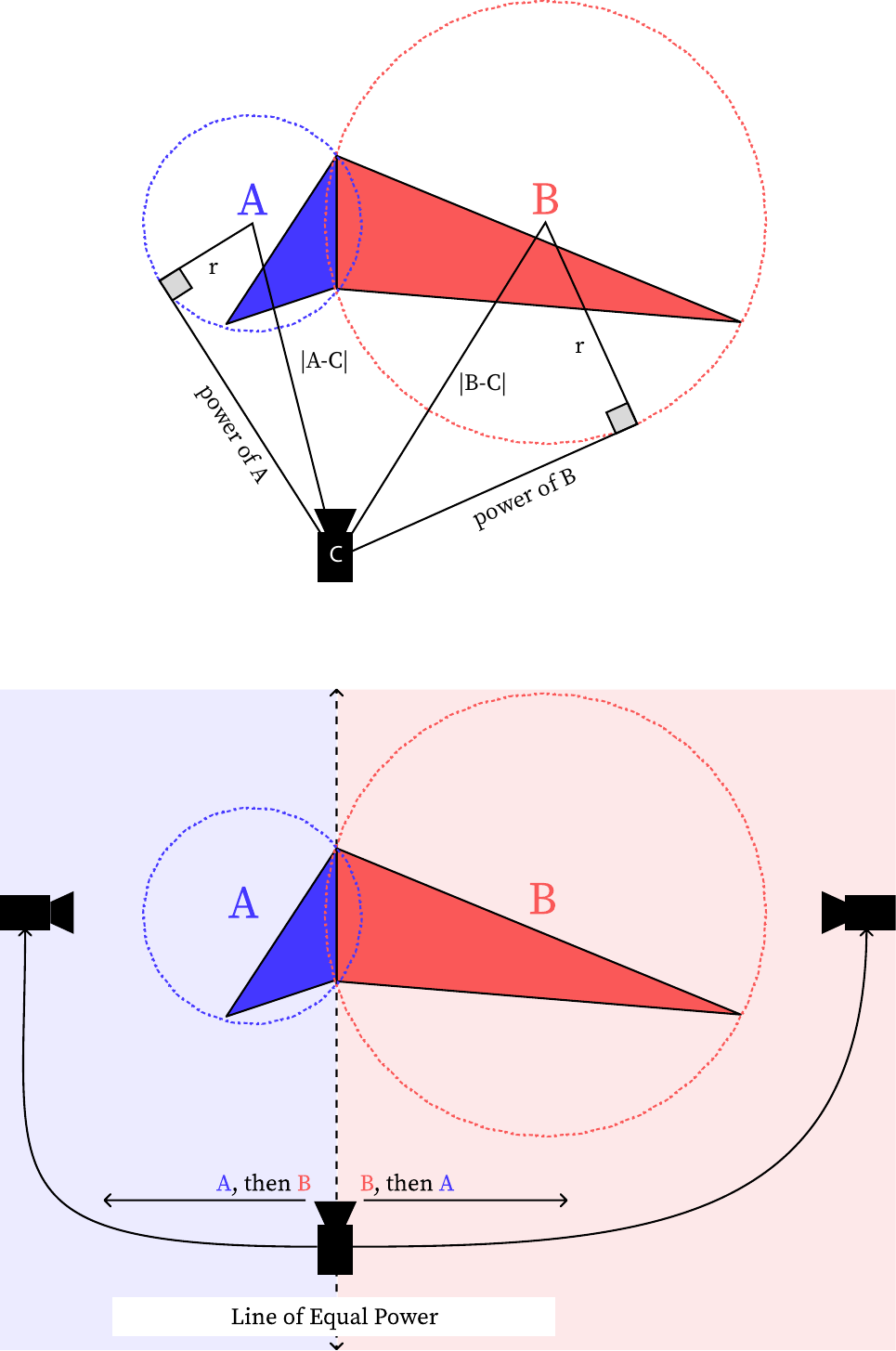}
    \caption{To sort two adjacent triangles relative to a viewpoint, we need only look at their circumcircles. The radical axis (the line along which the power of a spheres is equal) goes through the two intersection points between the circles, and divides the circumscribed triangles.
    Consider the centers of two circumcircles $A$ and $B$ and a viewpoint $C$ located along the radical axis.
    Since $|A-C|$ increases as the camera moves right and $|B-C|$ decreases, the power of B is less than the power of A to the right, and the reverse is true to the left. This extends to non-intersecting circles, and can be used for non adjacent triangles. Combining this observation with the the Delaunay empty sphere property allows us to sort all triangles.
    }
    \label{fig:radical_axis}
\end{figure}

% To show by induction how we can sort a Delaunay triangulation using powers of circumspheres, we sort two arbitrary adjacent triangles which are not contained within each other's circumspheres.
% If two triangles satisfy the empty circumsphere property, we can always depict them using two circles where one edge lines along the intersection between the two circles, and the rest of each triangle lies on the other side of their respective circle. The intuition here is that the line of equal power, or the radical axis, lies along this shared edge. This also happens to be the line along which the sort order changes. It is then trivial to show that the power of the further primitive increases relative to the closer one, thereby yielding the correct sort order for the two primitives. This proof generalizes to 3D.
To demonstrate how a Delaunay triangulation can be visibility-sorted relative to a specific viewpoint using the power of circumcircles, consider two adjacent triangles. Geometrically, the edge shared by these two triangles lies along the radical axis (the line of intersection) of their respective circumcircles.
This axis is the line of equal power; it divides the plane into two regions. On one side, the viewpoint has a lower power distance to the first circumcircle, but on the other side it has a lower power distance to the second. Since the relative depth order of the two triangles flips only when the viewpoint crosses this axis, the power distance serves as a consistent sorting criterion. This implies that sorting the primitives by the power of the viewpoint relative to their circumcircles yields the correct visibility order. This property naturally generalizes to 3D, where the shared face of two tetrahedra lies on the radical plane. See Figure~\ref{fig:radical_axis} for a visualization.

\section{Details}

% Combined with the viewing direction, this means we have our function $f_\theta$ has domain: circumcenter $\in R^3$, radius $\in R_{>0}$, viewing direction $\in R^3$. 
% Putting it all together, during optimization, the scene is represented by the following optimizable parameters: a set of 3D points $\mathcal{X}$ and a neural network weights $\theta$.
% The 3D Delaunay triangulation of these points yields a set of tetrahedra $\{T_k\}_{k=1}^N$.
% These tetrahedra can then be associated with a constant density and linearly varying color by querying $f_\theta$ at the circumcenter $\circumcenter_k$ and corresponding circumradius $\circumradius_k$, ith viewing direction $d_k = \frac{\centroid_k - o}{\|\centroid_k - o\|}$, where $\centroid_k$ is the centroid of tetrahedron $k$:
% \[
% \{\sigma_k, \colorc_k, \cgradient_k\} = f_\theta\left( \stopnabla{}  \circumcenter_k, \stopnabla{} \circumradius_k, \stopnabla{} d_k\right)
% \]
% where \(\sigma_k\in\mathbb R_{\ge0}\) is the constant density, \(\colorc_k\in\mathbb R^3\) is the base color centered at the circumsphere center, \(\cgradient\in\mathbb R^3\) is the monochromatic color gradient, and $\stopnabla{}$ is the stop gradient symbol.

% We stop the gradient across these inputs into $f$ to avoid determining the position of the vertices according to the linear interpolation gradients, which are noisy from the hash function.

Even though we are using GPU acceleration, recomputing a Delaunay triangular from scratch is still somewhat slow.
As such, we only update vertex locations and their corresponding Delaunay topology every 10 iterations.
On a scene with 1.2 million tetrahedra, our timings during training are approximately 19 milliseconds for rendering, 13 milliseconds to query the instant-NGP, 143 milliseconds for the backwards pass, and 92 milliseconds to compute the Delaunay triangulation.

Because densification changes the circumcenters of the primitives and therefore their colors, we temporarily increase the learning rate of some of the model to allow it to rapidly  accommodate for this change.
After densification at iteration $I$, we increase the learning rate of the instant NGP weights and the vertex coordinates
% $\theta$ and $\mathcal{X}$
by adding a ``spike'' function with duration $L$ on top of the decay function $l(i)$:
\begin{align}
    \mathbbm{1}\{i-I>0\}\left(l(0) - l(I)\right)\exp\left(-\frac{6(i-I)}{L}\right)
\end{align}
This increases the decaying learning rate back up to its maximum (initial) value, before decaying it back down in $L$ iterations.

\subsection{Algorithm Details}\label{sec:rasterization-details}
The naive version of our shader would, for each of the 4 vertices of each tetrahedra, load all of the tetrahedra information and attached it each of the vertices.
Then, when the fragment shader for any given triangle is called, the information from the closest vertex would be obtained, and the full Cyrus-Beck algorithm~\cite{cyrus1978} would be run on the rest of the tetrahedra to compute the back intersection, which then allows us to compute the integral along the intersection with the primitive.

In our optimized implementation, we leverage the capabilities of a mesh shader to allow the four vertices of each tetrahedra to share information between themselves. This reduces the number of loads by a factor of four. We also optimize the Cyrus-Beck algorithm: in the vertex shader we precompute the intersection distance with each of the vertices, and then perform interpolation between these distances to find the intersection distance with each plane. This interpolation must be perspective-correct barycentric interpolation to ensure that the interpolation results are correct. A similar technique can be applied to the color gradient to further simplify the fragment shader.
This reduces runtime cost to only 12 FLOPs. See Algorithm~\ref{alg:shader}.

\begin{algorithm}[htbp]
    \caption{Shader code to perform accelerated intersections using wave intrinsics within a mesh shader.}
    \label{alg:shader}

    % --- Block 1: Mesh Shader ---
    \begin{lstlisting}[title={Mesh shader intersection pre-calculation}]
uint bli = (WaveGetLaneIndex() / 4) * 4;
#define WRTV(x, id) WaveReadLaneAt(x, bli + id)

static const uint4 kTetTriangles[4] = {
    uint4(0, 2, 1, 3),
    uint4(1, 2, 3, 0),
    uint4(0, 3, 2, 1),
    uint4(3, 0, 1, 2),
};

tri = kTetTriangles[tetVertexId];

// outward facing normal
n = cross(
    WRTRV(vertex, tri[2]) -
    WRTRV(vertex, tri[0]),
    WRTRV(vertex, tri[1]) -
    WRTRV(vertex, tri[0]));

v = WRTRV(vertex, tri[0])
num = dot(n, v - rayOrigin);

o.planeN = float4(
    WRTRV(num, 0),
    WRTRV(num, 1),
    WRTRV(num, 2),
    WRTRV(num, 3) );
o.planeD = float4(
    dot(WRTRV(n, 0), o.rayDir),
    dot(WRTRV(n, 1), o.rayDir),
    dot(WRTRV(n, 2), o.rayDir),
    dot(WRTRV(n, 3), o.rayDir) );
return o;
    \end{lstlisting}

    % --- Block 2: Fragment Shader ---
    \begin{lstlisting}[title={Fragment shader intersection routine}]
d = length(rayDir);
o.planeD /= d;

all_t = o.planeN / o.planeD;

t_enter = max(o.planeD > 0.0f ? all_t : -FLT_MAX);
t_exit  = min(o.planeD < 0.0f ? all_t : FLT_MAX);
    \end{lstlisting}

\end{algorithm}

% \begin{algorithm}[htbp]
%     \caption{Shader code to perform accelerated intersections using wave intrinsics within a mesh shader.
%     }
%     \label{alg:shader}
% \begin{minted}
% [frame=lines,
%  framerule=1pt,
%  % rulecolor=\color{black!20},
%  label={Mesh shader intersection pre-calculation},
%  fontsize=\footnotesize]
% {hlsl}
% uint bli = (WaveGetLaneIndex() / 4) * 4;
% #define WRTV(x, id) WaveReadLaneAt(x, bli + id)

% static const uint4 kTetTriangles[4] = {
%     uint4(0, 2, 1, 3),
%     uint4(1, 2, 3, 0),
%     uint4(0, 3, 2, 1),
%     uint4(3, 0, 1, 2),
% };

% tri = kTetTriangles[tetVertexId];

% // outward facing normal
% n = cross(
%     WRTRV(vertex, tri[2]) -
%     WRTRV(vertex, tri[0]),
%     WRTRV(vertex, tri[1]) -
%     WRTRV(vertex, tri[0]));

% v = WRTRV(vertex, tri[0])
% num = dot(n, v - rayOrigin);

% o.planeN = float4(
%     WRTRV(num, 0),
%     WRTRV(num, 1),
%     WRTRV(num, 2),
%     WRTRV(num, 3) );
% o.planeD = float4(
%     dot(WRTRV(n, 0), o.rayDir),
%     dot(WRTRV(n, 1), o.rayDir),
%     dot(WRTRV(n, 2), o.rayDir),
%     dot(WRTRV(n, 3), o.rayDir) );
% return o;
% \end{minted}

% \begin{minted}
% [frame=lines,
%  framerule=1pt,
%  % rulecolor=\color{black!20},
%  label={Fragment shader intersection routine},
%  fontsize=\footnotesize]
% {hlsl}
% d = length(rayDir);
% o.planeD /= d;

% all_t = o.planeN / o.planeD;

% t_enter = max(o.planeD > 0.0f ? all_t : -FLT_MAX);
% t_exit  = min(o.planeD < 0.0f ? all_t : FLT_MAX);
% \end{minted}
% % \end{figure}
% \end{algorithm}

\section{Results}

See Figure~\ref{fig:comparison_supp} for additional results.

\newcommand{\graphimone}[1]{
\begin{tikzpicture}[zoomboxarray, zoomboxarray rows=1, zoomboxes below, zoomboxarray inner gap=0.0cm]
    \node [image node] { \includegraphics[width=3.5cm,valign=b]{#1} };
    \zoombox[magnification=3]{0.25,0.18}
\end{tikzpicture}
}

\newcommand{\graphimtwo}[1]{
\begin{tikzpicture}[zoomboxarray, zoomboxarray rows=1, zoomboxes below, zoomboxarray inner gap=0.0cm]
    \node [image node] { \includegraphics[width=3.5cm,valign=b]{#1} };
    \zoombox[magnification=5]{0.35,0.87}
\end{tikzpicture}
}

\newcommand{\graphimthree}[1]{
\begin{tikzpicture}[zoomboxarray, zoomboxarray rows=1, zoomboxes below, zoomboxarray inner gap=0.0cm]
    \node [image node] { \includegraphics[width=3.5cm,valign=b]{#1} };
    \zoombox[magnification=4]{0.80,0.55}
\end{tikzpicture}
}

\newcommand{\graphimfour}[1]{
\begin{tikzpicture}[zoomboxarray, zoomboxarray rows=1, zoomboxes below, zoomboxarray inner gap=0.0cm]
    \node [image node] { \includegraphics[width=3.5cm,valign=b]{#1} };
    \zoombox[magnification=3]{0.45,0.36}
\end{tikzpicture}
}

\begin{figure*}[ht]
\centering
\bgroup
\begin{tabular}{@{}c@{}c@{}c@{}c@{}c@{}}
\graphimone{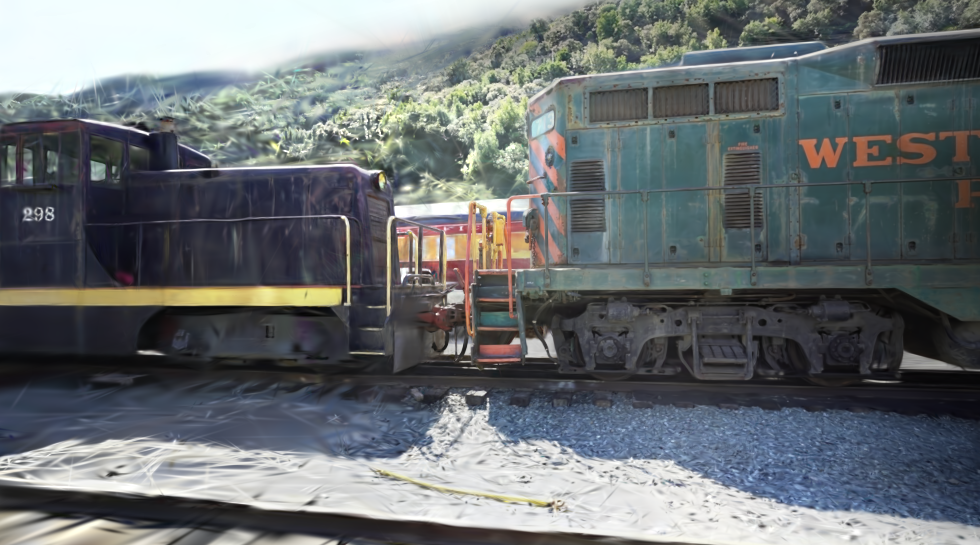} &
\graphimone{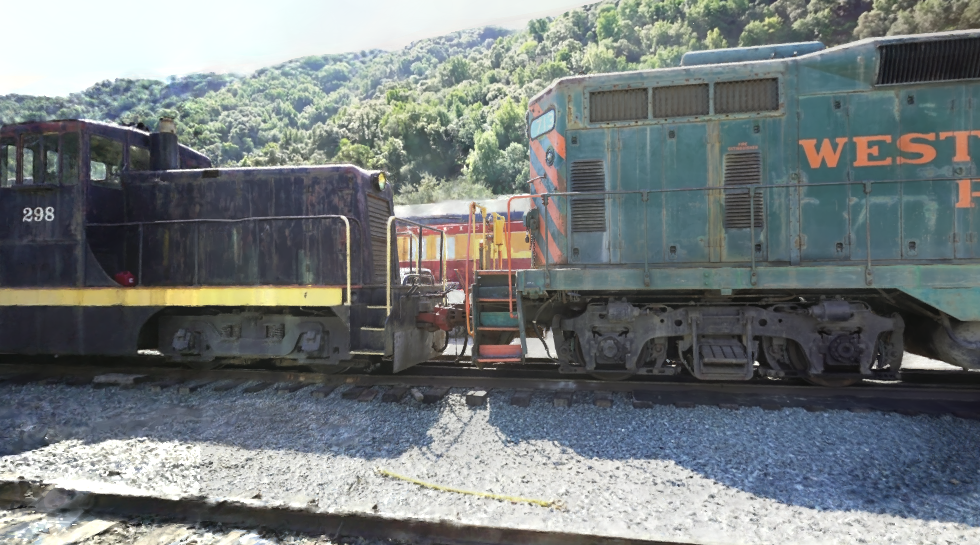} &
\graphimone{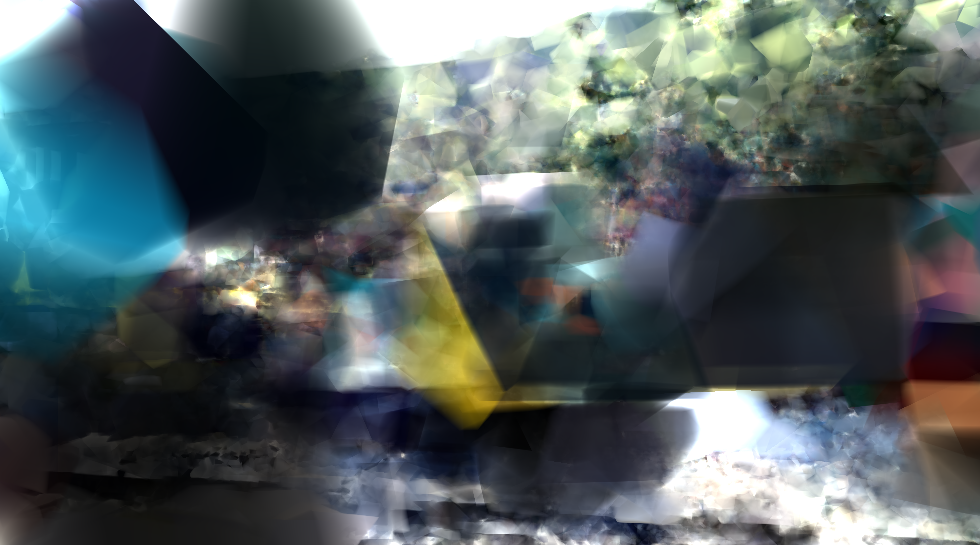} &
\graphimone{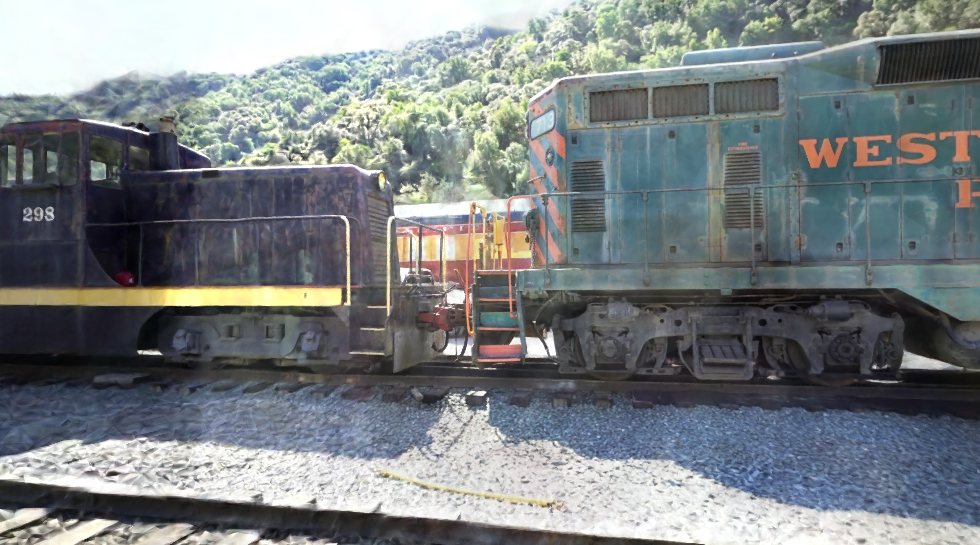} &
\graphimone{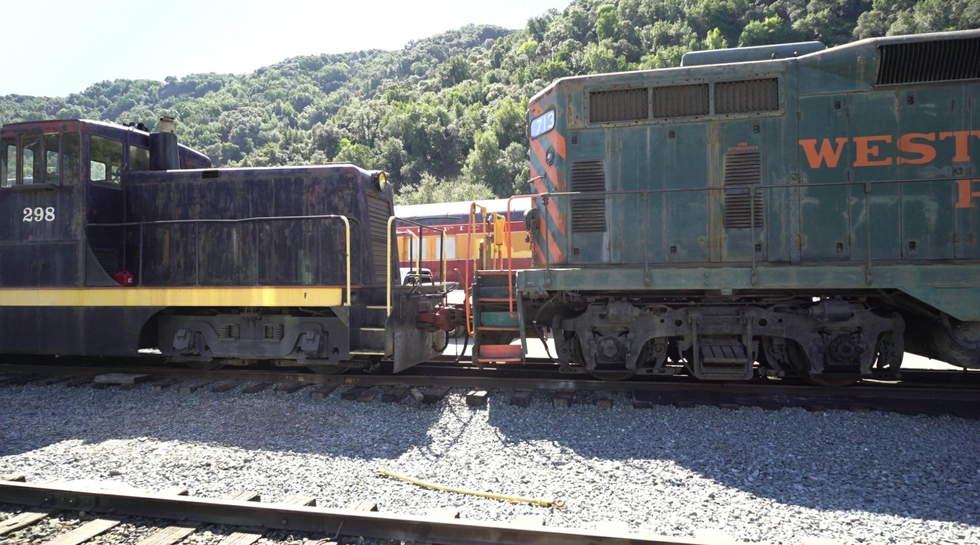} \\
\graphimtwo{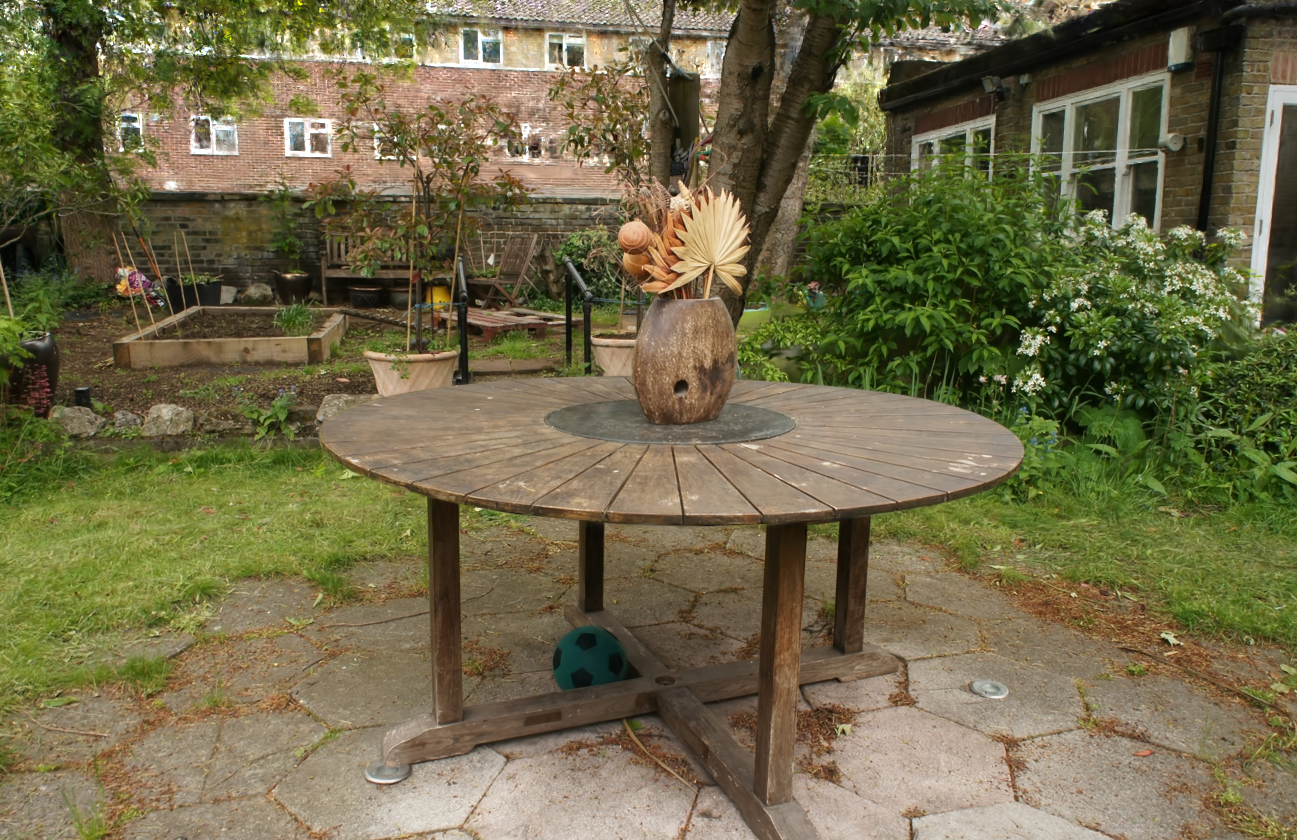} &
\graphimtwo{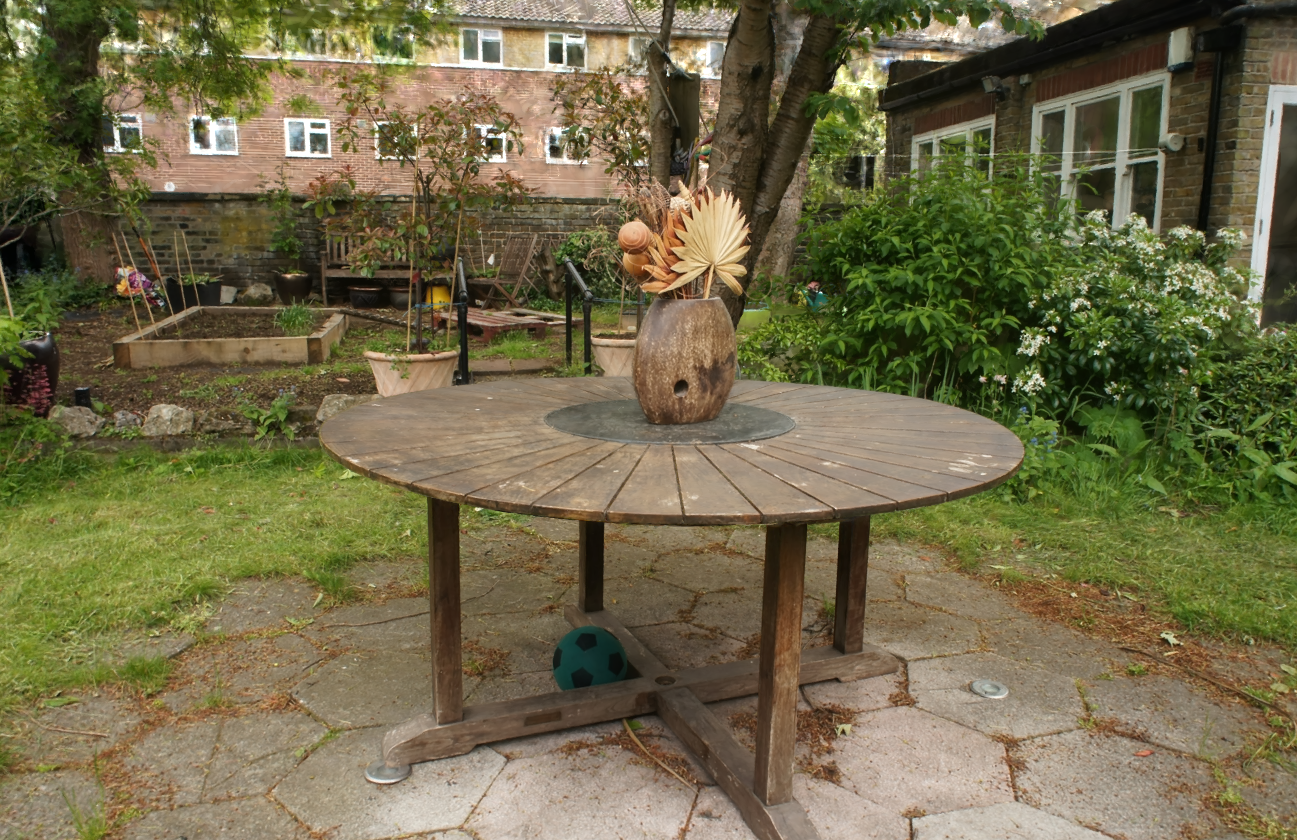} &
\graphimtwo{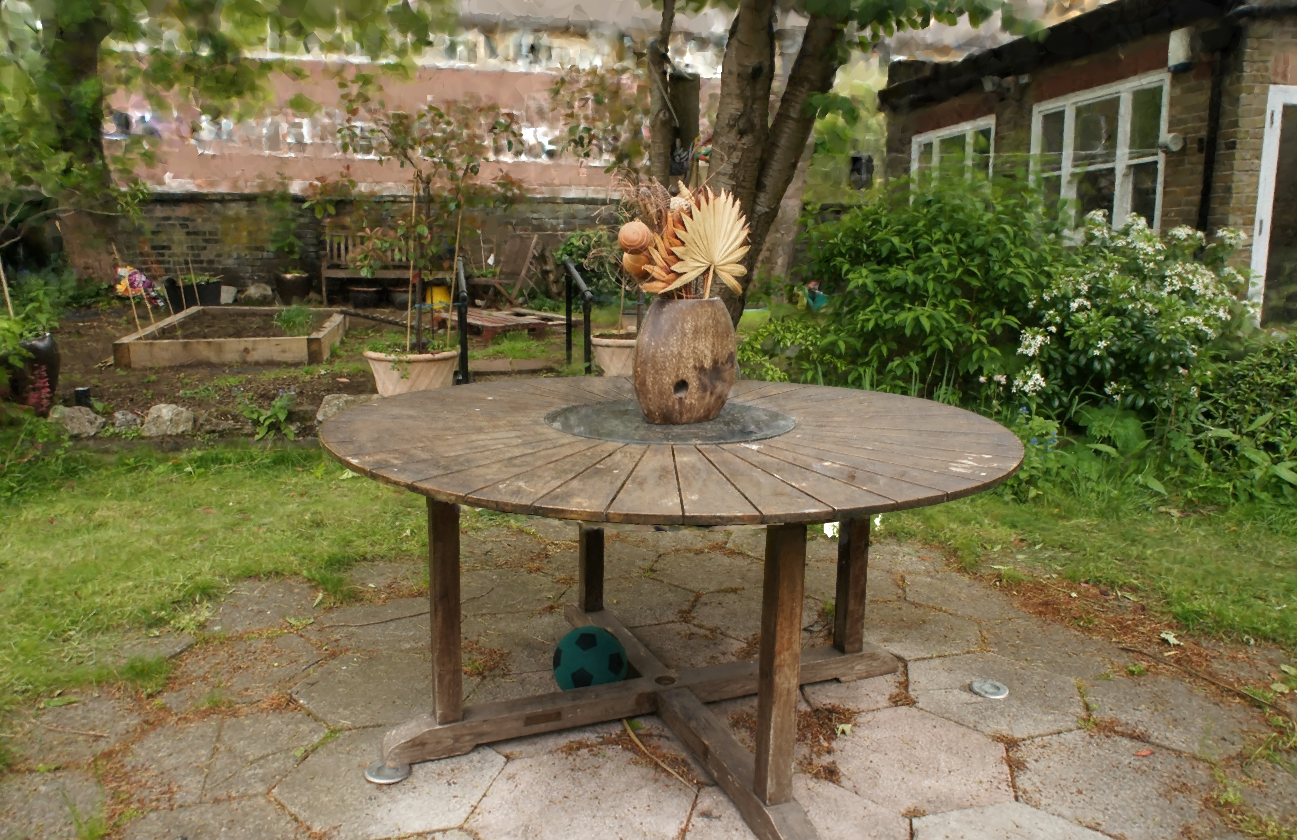} &
\graphimtwo{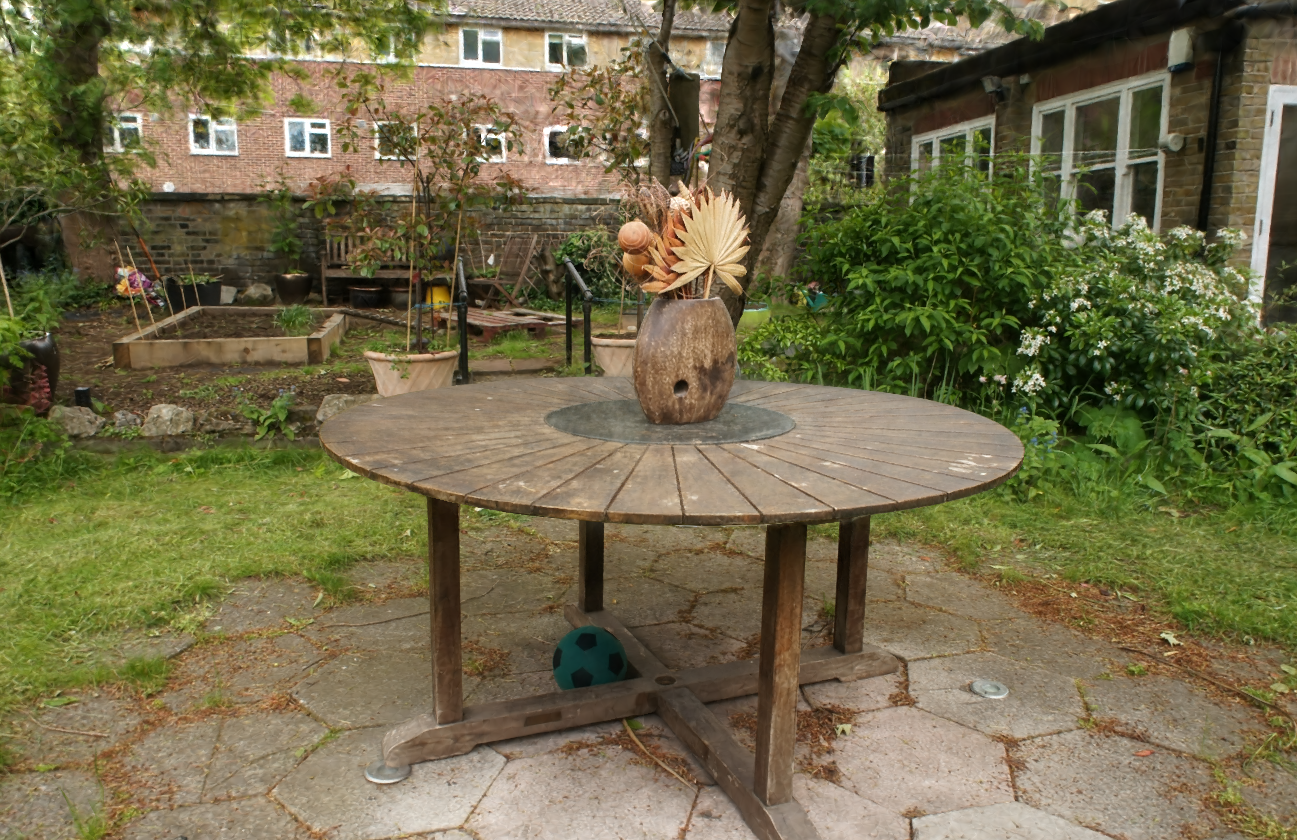} &
\graphimtwo{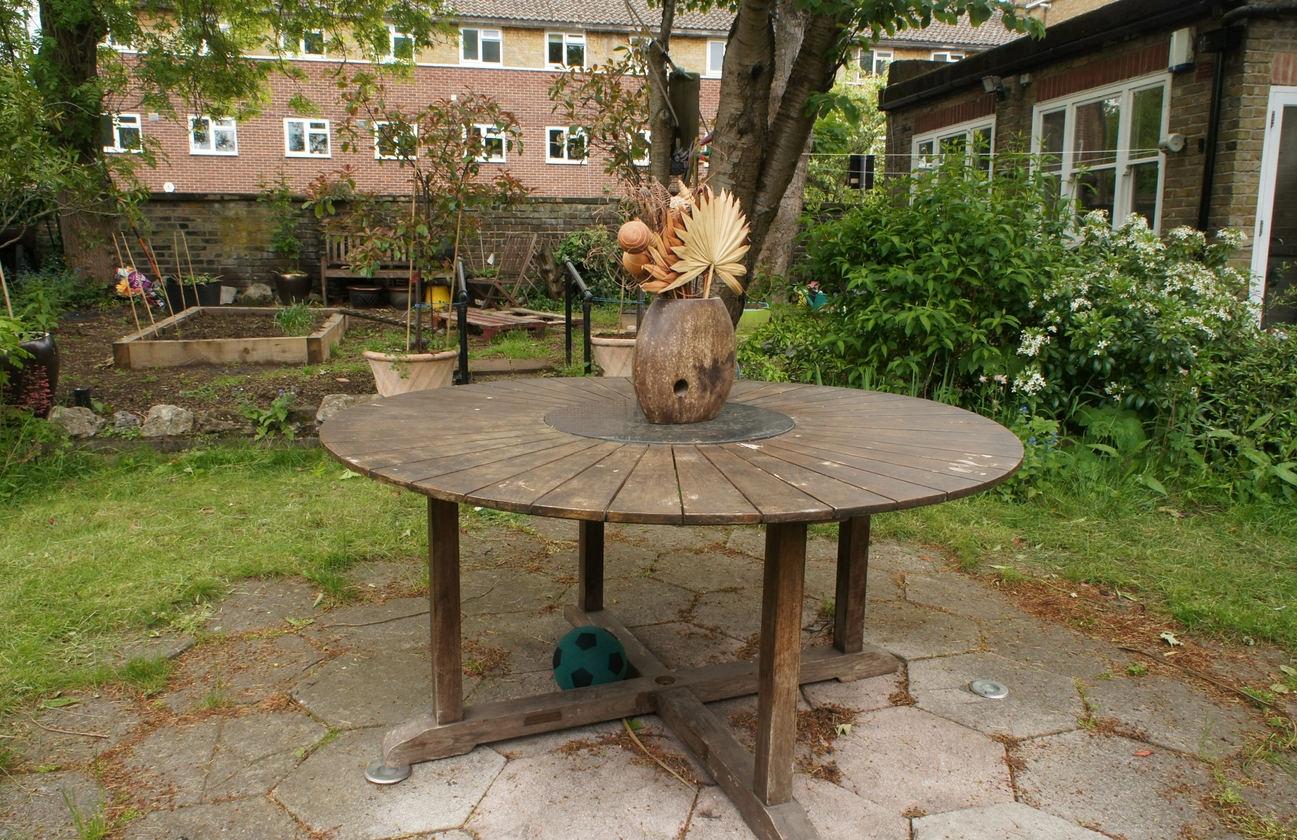} \\
\graphimthree{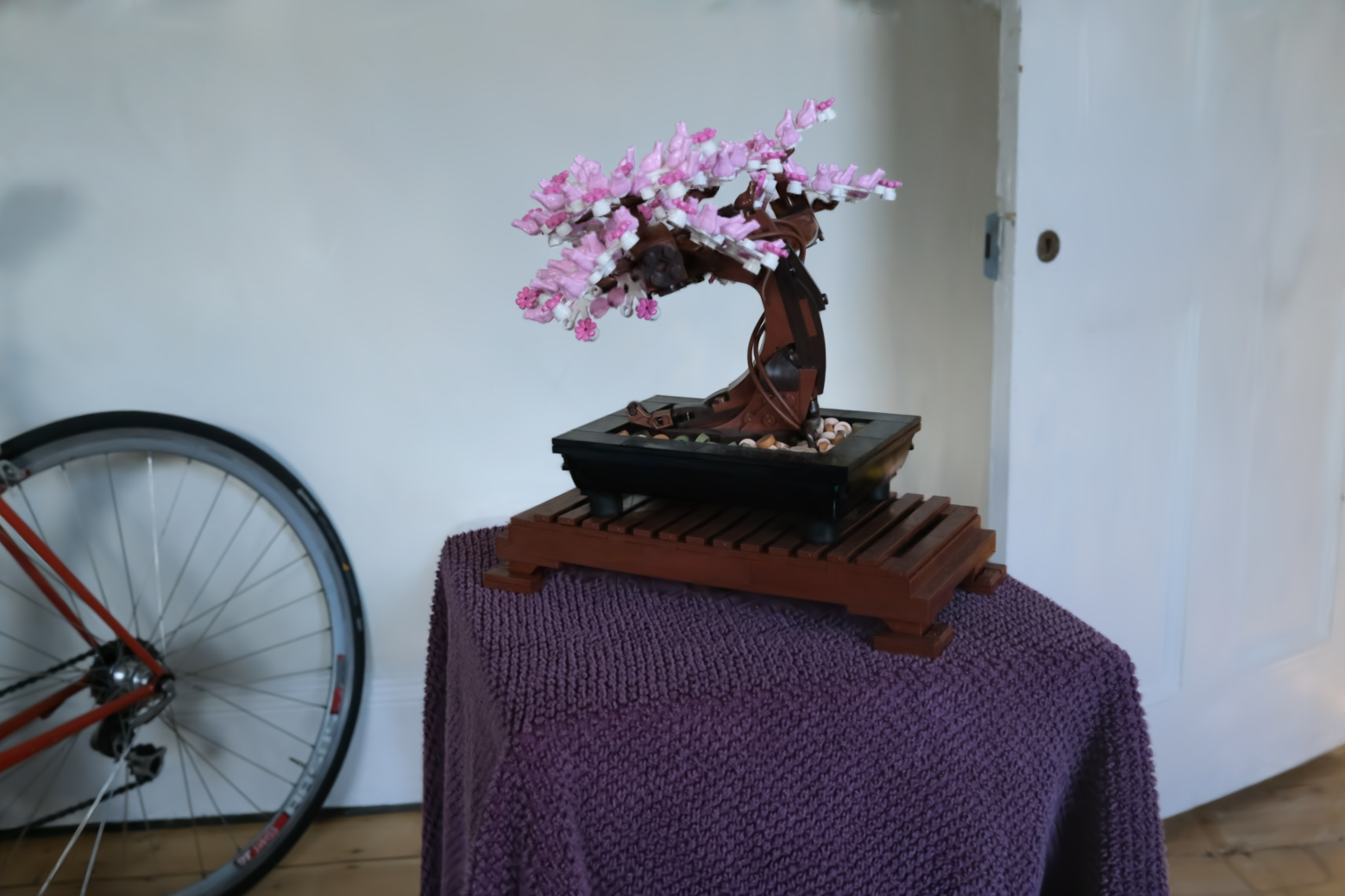} &
\graphimthree{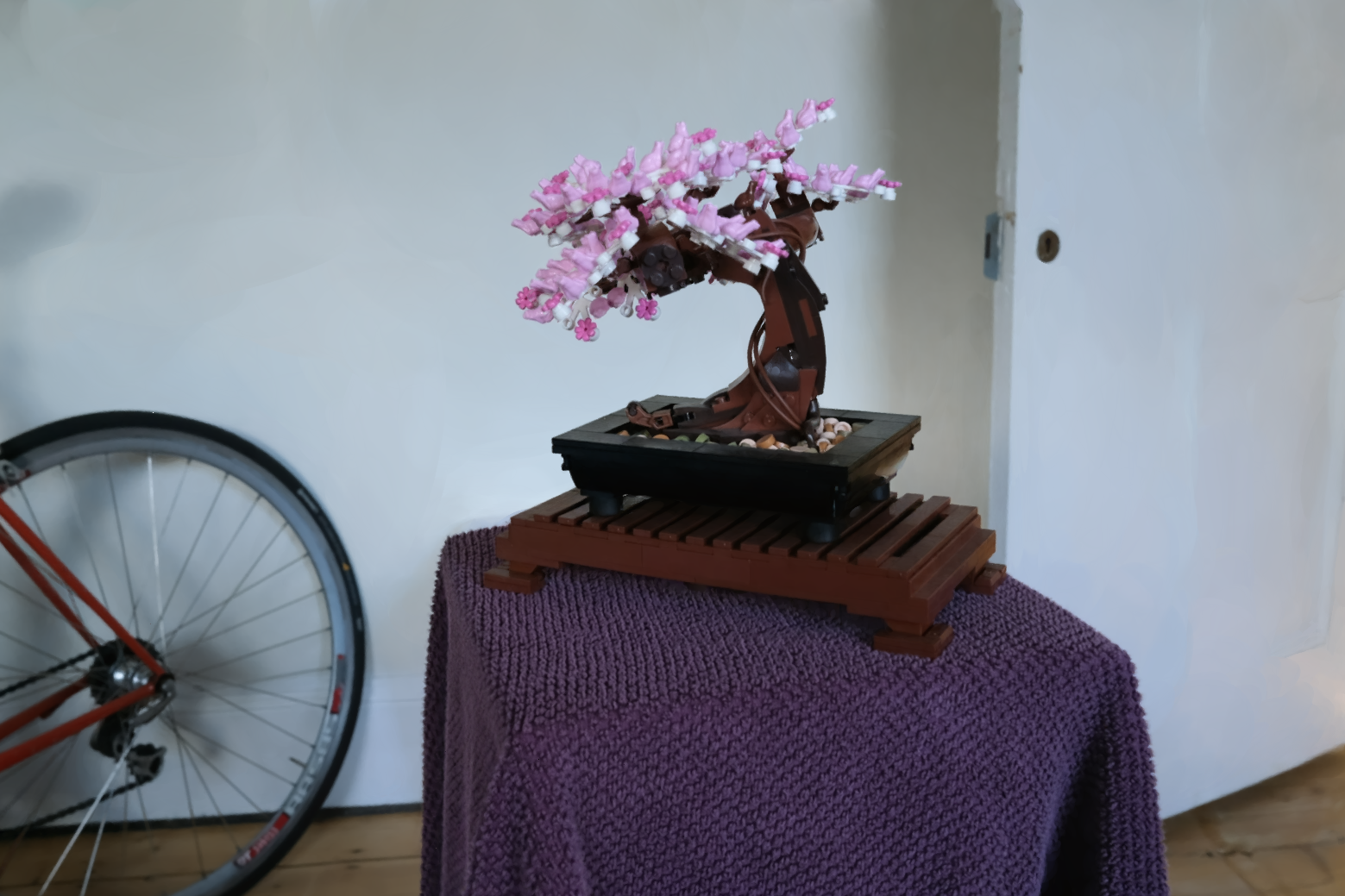} &
\graphimthree{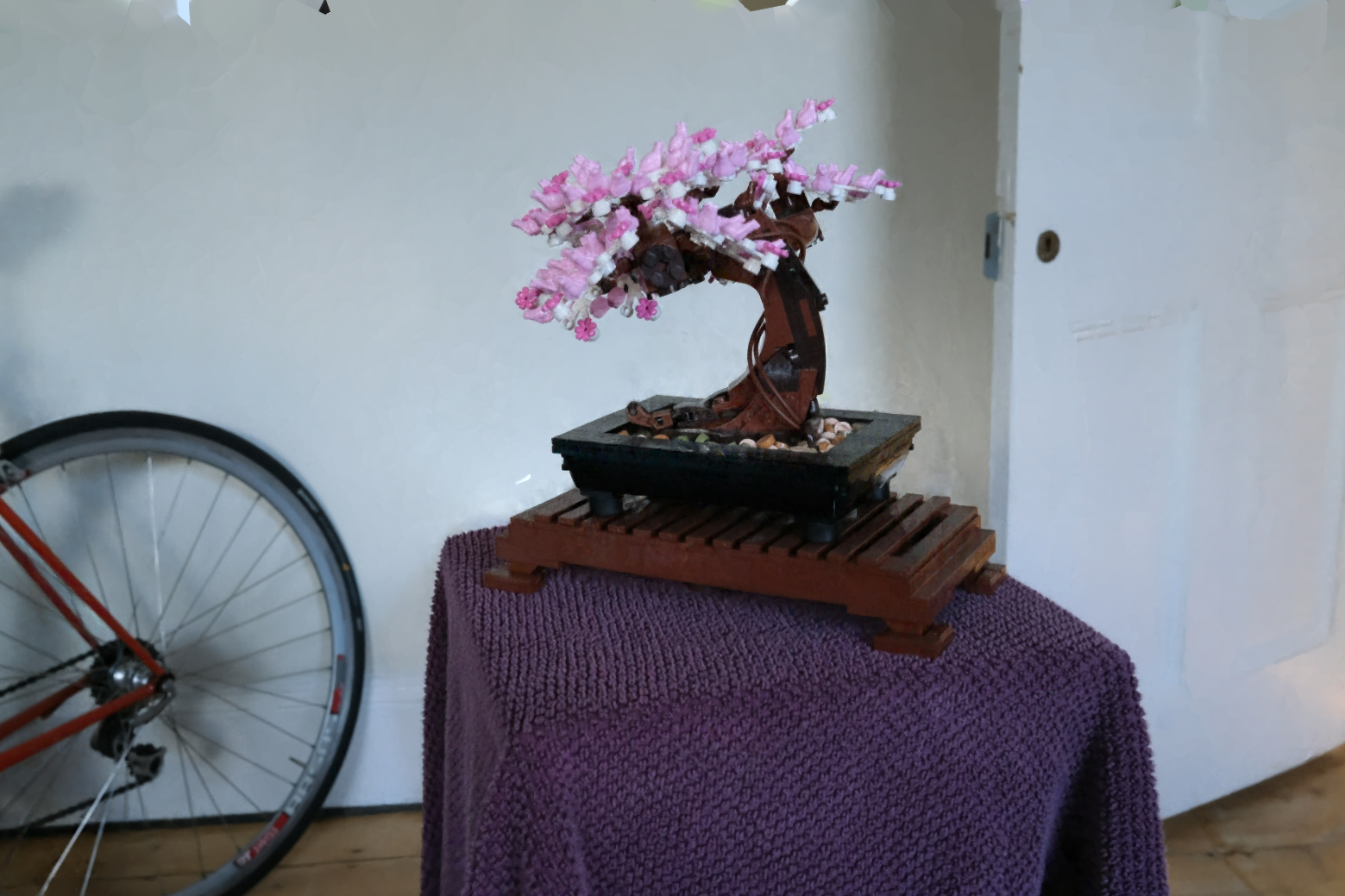} &
\graphimthree{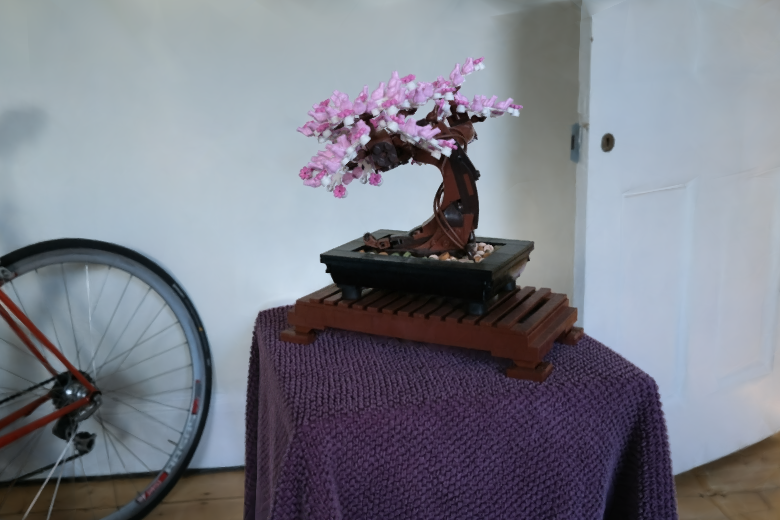} &
\graphimthree{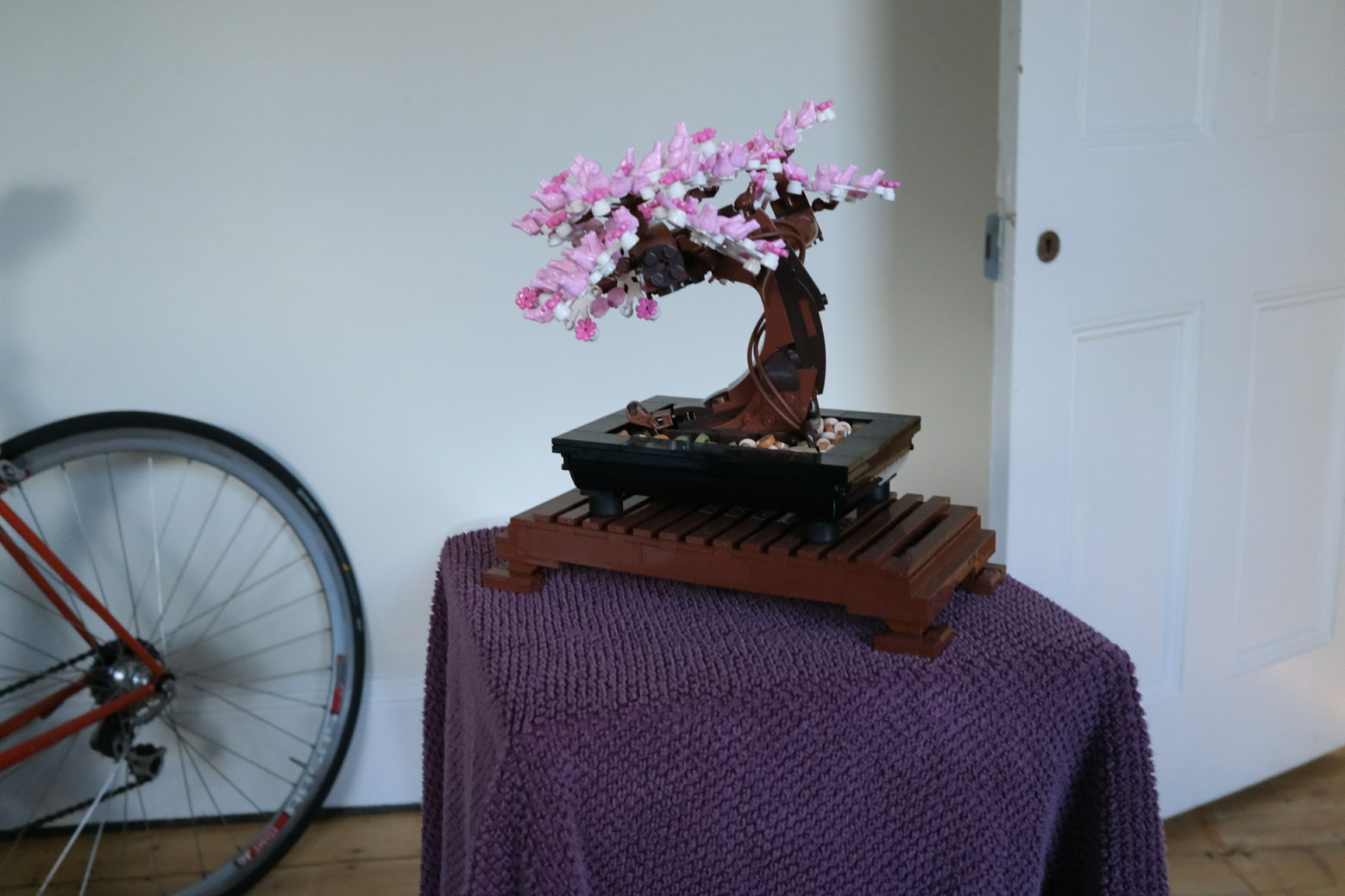} \\
\graphimfour{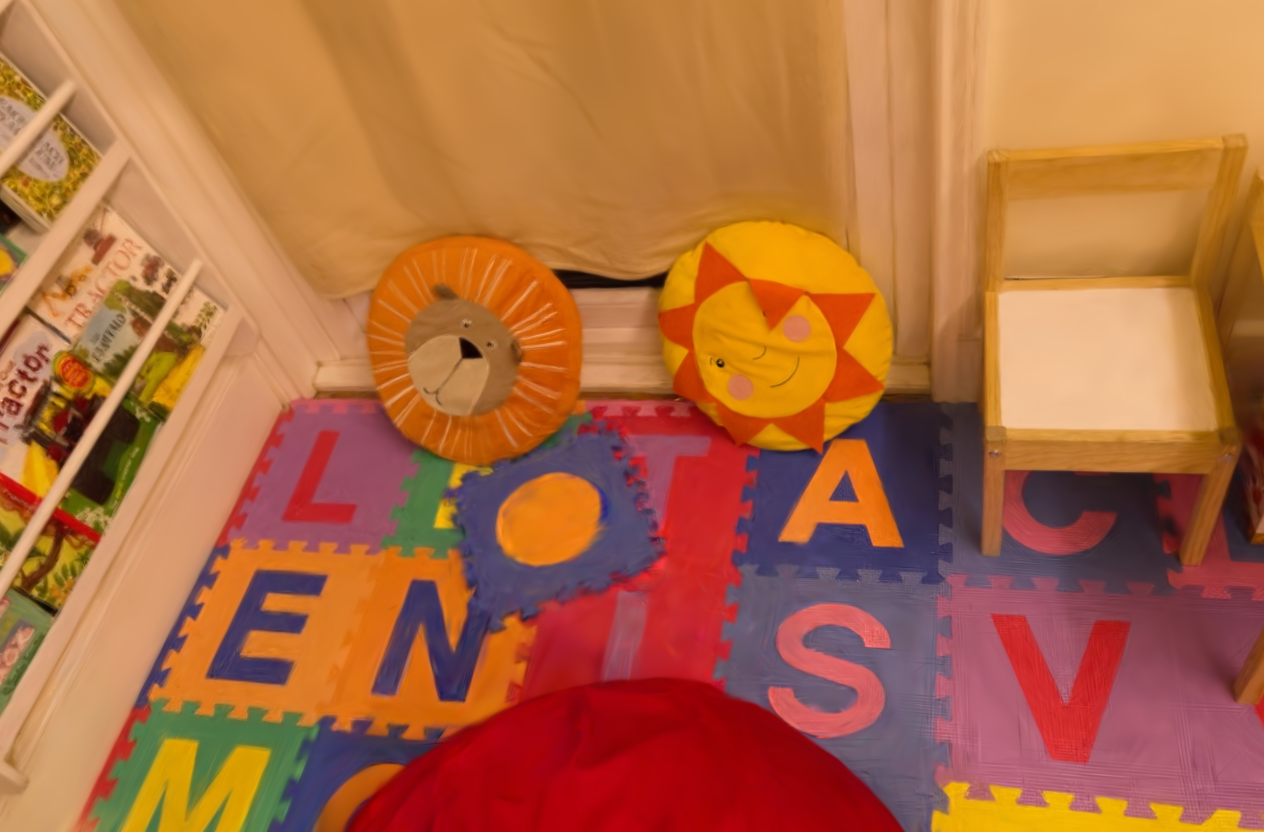} &
\graphimfour{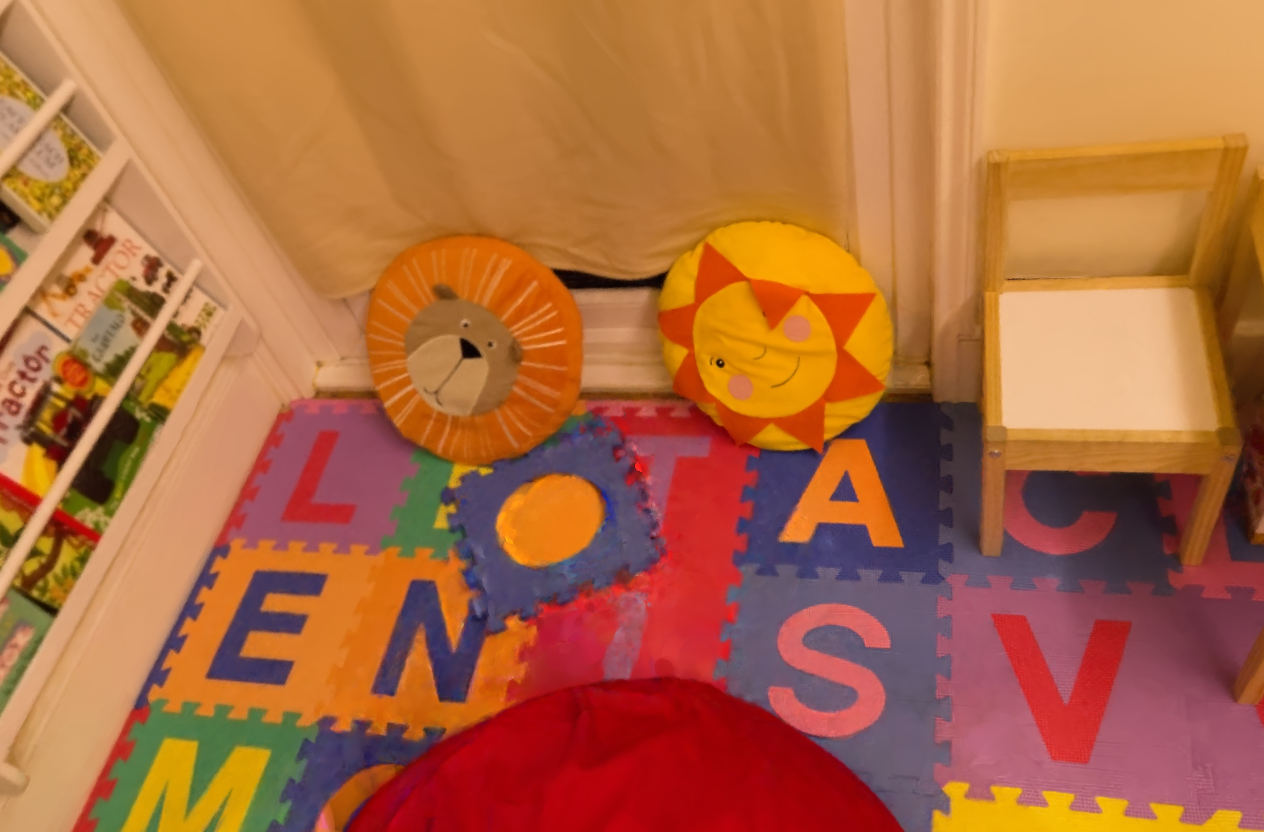} &
\graphimfour{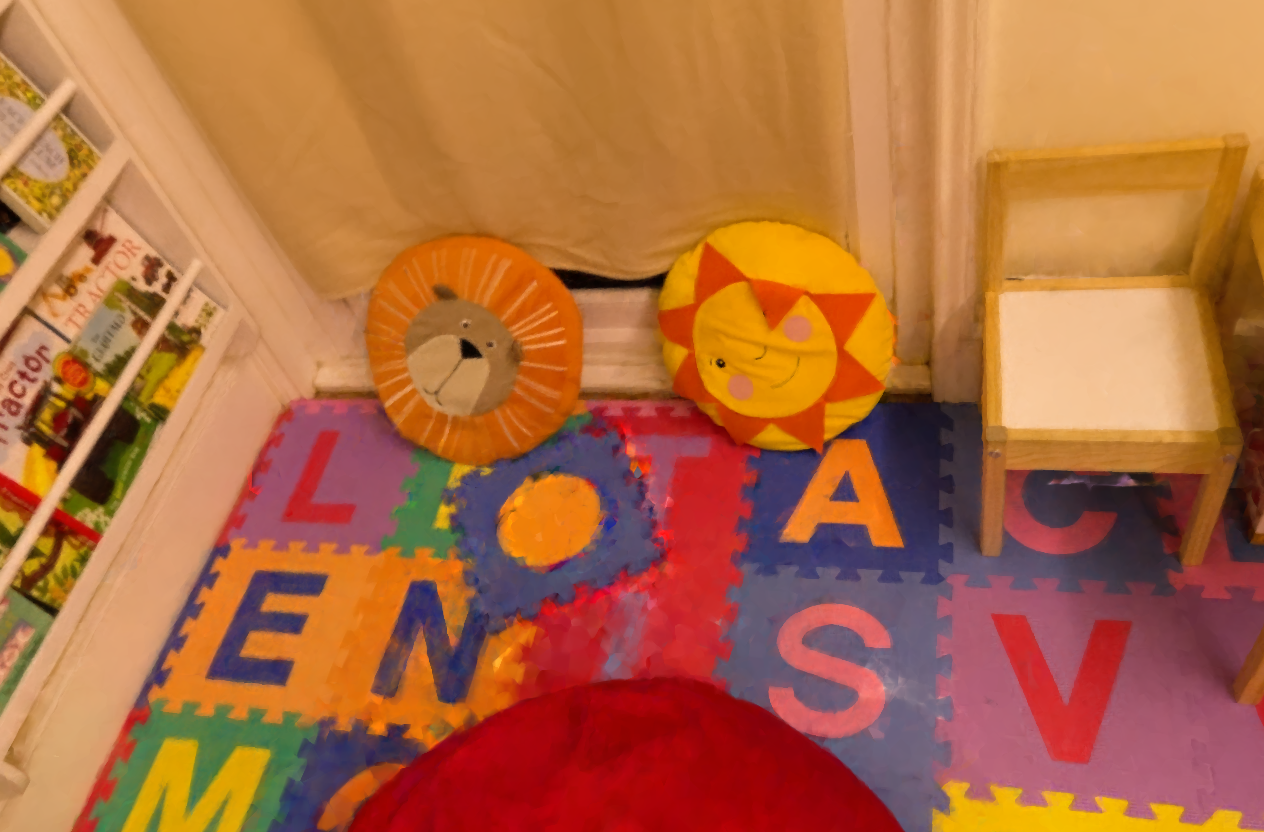} &
\graphimfour{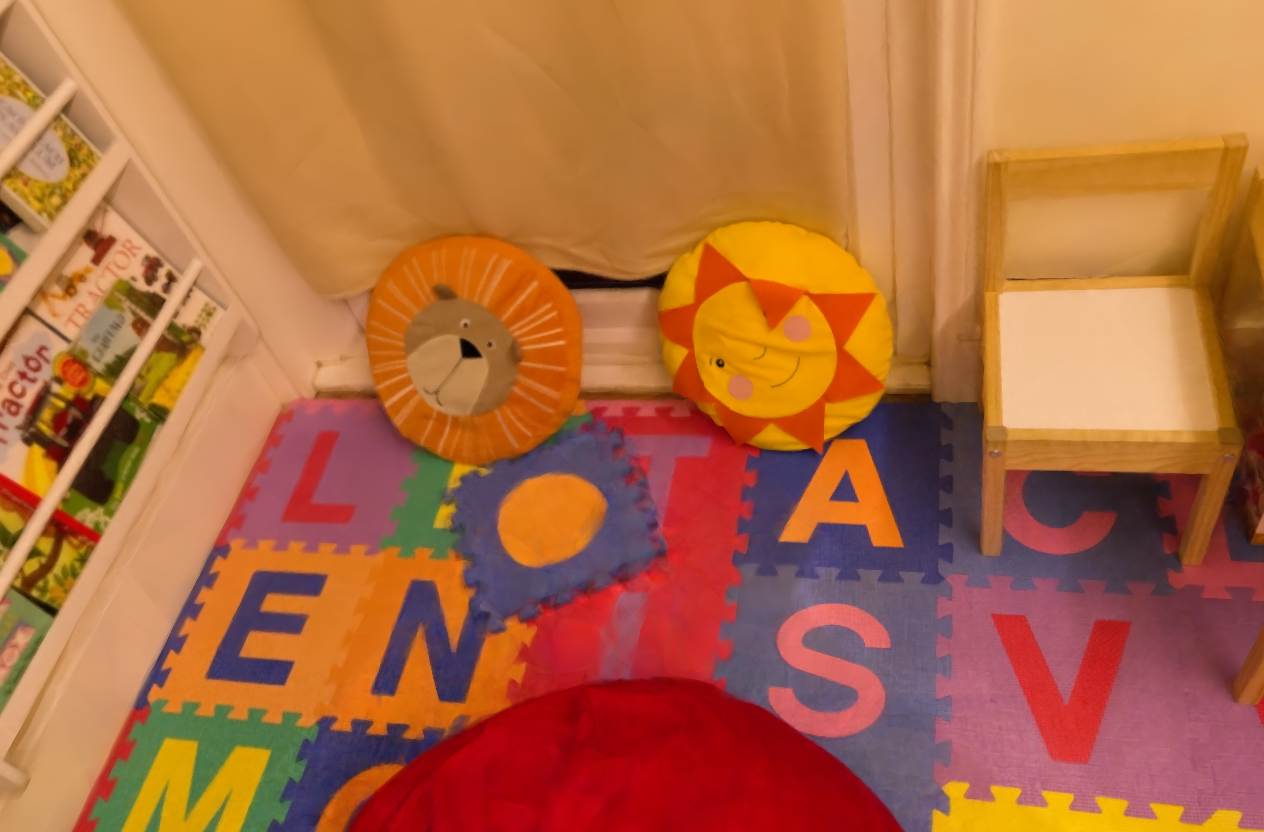} &
\graphimfour{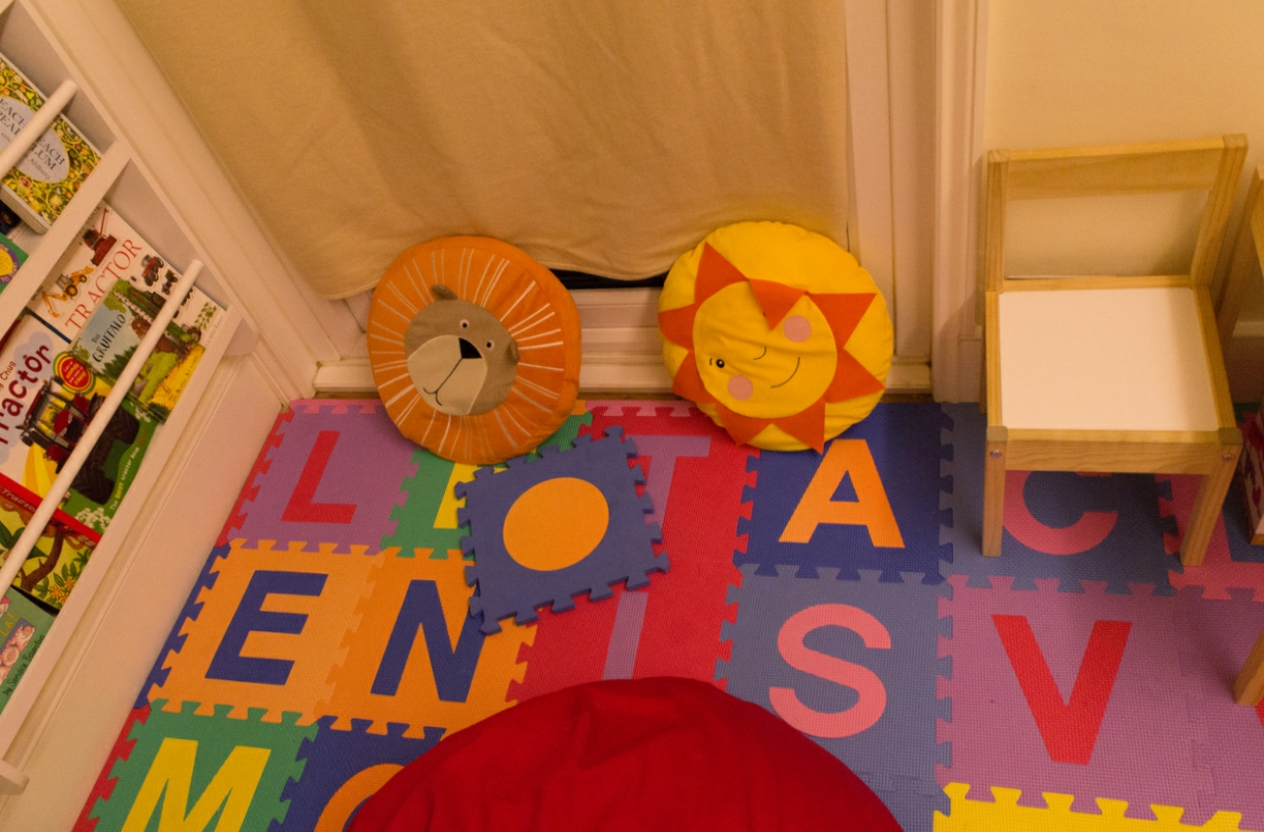} \\
3DGS & EVER & RadiantFoam & Ours & GT
\end{tabular}
\egroup
\caption{Visual comparison of our method alongside baseline algorithms on various scenes from the datasets we use.}
\label{fig:comparison_supp}
\end{figure*}

\end{document}